\documentclass[preprint,showpacs,preprintnumbers,amsmath,amssymb]{revtex4}

\usepackage{graphicx}
\usepackage{dcolumn}
\usepackage{bm}
\usepackage{hyperref}
\hypersetup{colorlinks=false}

\begin{document}

\newcommand{\RR}[1]{[#1]}
\newcommand{\intsum}{\sum \kern -15pt \int}
\newfont{\Yfont}{cmti10 scaled 2074}
\newcommand{\Y}{\hbox{{\Yfont y}\phantom.}}
\def\O{{\cal O}}
\newcommand{\bra}[1]{\left< #1 \right| }
\newcommand{\braa}[1]{\left. \left< #1 \right| \right| }
\def\Bra#1#2{{\mbox{\vphantom{$\left< #2 \right|$}}}_{#1}
\kern -2.5pt \left< #2 \right| }
\def\Braa#1#2{{\mbox{\vphantom{$\left< #2 \right|$}}}_{#1}
\kern -2.5pt \left. \left< #2 \right| \right| }
\newcommand{\ket}[1]{\left| #1 \right> }
\newcommand{\kett}[1]{\left| \left| #1 \right> \right.}
\newcommand{\scal}[2]{\left< #1 \left| \mbox{\vphantom{$\left< #1 #2 \right|$}}
\right. #2 \right> }
\def\Scal#1#2#3{{\mbox{\vphantom{$\left<#2#3\right|$}}}_{#1}
{\left< #2 \left| \mbox{\vphantom{$\left<#2#3\right|$}} \right. #3
\right> }}


\title{Three-Nucleon Bound State in a Spin-Isospin Dependent\\ Three Dimensional Approach}

\author{S. Bayegan}%
 \email{bayegan@khayam.ut.ac.ir}
 \author{M.~R. Hadizadeh}
\email{hadizade@khayam.ut.ac.ir}
\author{M. Harzchi}
\email{harzchi@ut.ac.ir}

\affiliation{ Department of Physics, University of Tehran, P.O.Box
14395-547, Tehran, Iran
}%

\date{\today}

\begin{abstract}
A spin-isospin dependent Three-Dimensional approach based on
momentum vectors for formulation of the three-nucleon bound state is
presented in this paper. The three-nucleon Faddeev equations with
two-nucleon interactions are formulated as a function of vector
Jacobi momenta, specifically the magnitudes of the momenta and the
angle between them with the inclusion of the spin-isospin quantum
numbers, without employing a partial wave decomposition. As an
application the spin-isospin dependent Faddeev integral equations
are solved with Bonn-B potential. Our result for the Triton binding
energy with the value of $-8.152$ MeV is in good agreement with the
achievements of the other partial wave based methods.

\end{abstract}

\pacs{21.45.+v, 21.30.-x, 21.10.Dr, 27.10.+h, 21.10.Hw }
\keywords{Suggested keywords}
\maketitle

\section{Introduction} \label{sec:introduction}
During the past years, several methods have been developed to solve
the nonrelativistic Schr\"{o}dinger equation accurately for
few-nucleon bound states, by using realistic nuclear potentials.
These methods are the CRCGV \cite{Hiyama-PRL85}, the SV
\cite{Usukura-PRB59}, the HH \cite{Viviani-PRC71}, the GFMC
\cite{Viringa-PRC62}, the NCSM \cite{Navratil-PRC62}, the EIHH
\cite{Barnea-PRC67} and the Faddeev. These calculational approaches
are mostly based on a partial wave (PW) decomposition. Stochastic
and Monte Carlo methods, however, are performed directly using the
position vectors in the configuration space. One of the most viable
approaches appears to be the Faddeev method.

The calculations based on the Faddeev approach are performed after a
PW expansion with phenomenological potentials either in the momentum
space \cite{Sammarruca-PRC46}-\cite{Fachruddin-PRC69} or in the
configuration space \cite{Chen-PRL55}- \cite{Friar-PLB311}. Recent
bound state calculations with the Faddeev approach have been done
with the chiral potentials in the momentum space
\cite{Bedaque-NPA676}-\cite{Platter-PLB607}. Experience in
three-nucleon calculations shows that the standard treatment based
on a PW decomposition is quite successful but also rather complex,
since each building block related to involved operators requires
extended algebra. The Faddeev calculations based on a PW
decomposition, which includes the spin-isospin degrees of freedom,
after truncation leads to a set of a finite number of coupled
equations in two variables for the amplitudes and one needs a large
number of partial waves to get converged results. In view of this
large number of interfering terms it appears natural to give up such
an expansion and work directly with the vector variables. On this
basis three- and four-body bound states have recently been studied
in a Three-Dimensional (3D) approach where the spin-isospin degrees
of freedom have been neglected in the first attempt
\cite{Elster-FBS27}-\cite{Hadizadeh-EPJA}. In the case of three-body
bound state the Faddeev equations have been formulated for three
identical bosons as a function of vector Jacobi momenta, with the
specific stress upon the magnitudes of the momenta and the angle
between them. Adding the spin-isospin to the 3D formalism is a major
additional task, which will increase more degrees of freedom into
the states and therefore will lead to a strictly finite number of
coupled equations \cite{Bayegan-EFB20}. In this paper we have
attempted to implement this task by including the spin-isospin
degrees of freedom in the 3N bound state formalism. To this end we
have formulated the Faddeev equations for the 3N bound state with
the advantage of using the realistic NN forces. The presented 3D
formalism in this paper in comparison with the traditional PW
formalism avoids the highly involved angular momentum algebra
occurring for the permutation operators. According to the
spin-isospin states that have been taken into account, we have
obtained the eight, twelve, sixteen and twenty four coupled
equations for a description of the 3N bound state, i.e. $^{3}H$ and
$^{3}He$. In this way, we solve the Faddeev integral equations for
calculation of the Triton binding energy with Bonn-B potential. The
input to our calculations is the two-body $t$-matrix which has been
calculated in an approach based on a Helicity representation and
depends on the magnitudes of the initial and final momenta and the
angle between them \cite{Fachruddin-PRC62}.

 This manuscript is organized as follows. In section
\ref{sec:formulation} we present the formalism. Meaning that we have
derived the Faddeev equations and the 3N wave function in a
realistic 3D scheme both as a function of Jacobi momenta vectors and
the spin-isospin quantum numbers. Also the novel 3D representation
of the Faddeev equations is contrasted with the corresponding
traditional PW representation. In section \ref{sec:numerical
results} we present our results for the Triton binding energy and
compare them with the results obtained from the PW calculations. In
order to test our calculations the calculated expectation values of
the Hamiltonian operator are compared to the obtained eigenvalue
energies. Finally in section \ref{sec:summary} a summary and an
outlook will be presented.

\section{Formulation For 3N Bound State in a 3D Faddeev Scheme } \label{sec:formulation}

\subsection{The Faddeev Equations} \label{subsec:Faddeev-equations}
\label{section II} The bound state of three pairwise-interacting
nucleons is described by the Faddeev equation \cite{Stadler-PRL78}:
\begin{eqnarray}
|\psi\rangle &\equiv& |\psi_{12,3}\rangle =G_{0}tP |\psi\rangle,
\label{eq.FC}
\end{eqnarray}
where $G_{0}$ is the free 3N propagator, $t$ denotes the NN
transition matrix determined by a two-body Lippman-Schwinger
equation and $P=P_{12}P_{23}+P_{13}P_{23}$ is the sum of a cyclic
and anti-cyclic permutations of the three nucleons. The total 3N
wave function $|\Psi\rangle$ is composed of the three Faddeev
components as:
\begin{equation}
|\Psi\rangle=(1+P)|\psi\rangle. \label{eq.WF}
\end{equation}

\begin{figure}[hbt]
\includegraphics*[width=14cm]{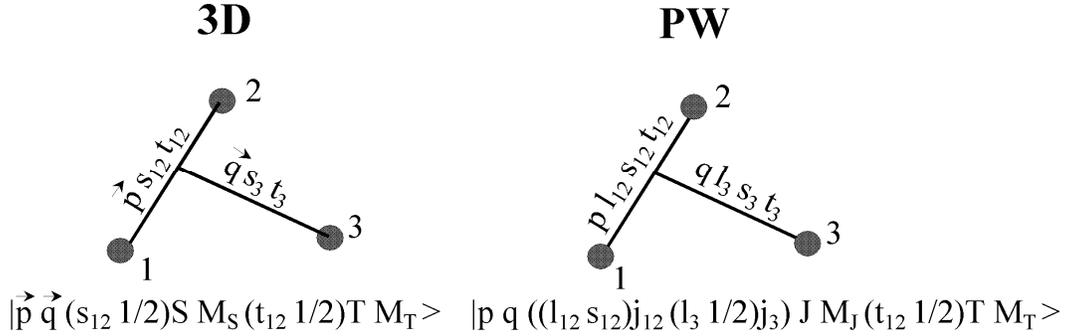}
\caption{\label{fig.basis_states} Definition of the 3N basis states
in the 3D approach in comparison with the corresponding basis states
in the PW approach.}
\end{figure}

The antisymmetry property of $|\psi\rangle$ under exchange of the
interacting particles $1$ and $2$ guarantees that $|\Psi\rangle$ is
totally antisymmetric. In order to solve Eq. (\ref{eq.FC}) in the
momentum space we introduce the 3N basis states in a 3D formalism
as, (see Fig.~\ref{fig.basis_states}):
\begin{eqnarray}
  |\, {\bf p} \, {\bf q} \,\,  \alpha  \, \rangle &\equiv& | \,
 {\bf p} \, {\bf q} \,\, \alpha_{S} \, \alpha_{T}\,  \rangle,
 \label{eq.basis}
\end{eqnarray}
the basis states involve two standard Jacobi momenta ${\bf p}$ and
${\bf q}$ \cite{Stadler-PRL78}, and $| \, \alpha \, \rangle$ is the
spin-isospin parts of the basis states, where the spin part is
defined as:
\begin{eqnarray}
  | \, \alpha_{S} \, \rangle &\equiv& | \, ((s_{1}\,\,s_{2})s_{12}
 \,\, s_{3}) S \, M_{S} \, \rangle \equiv | \, (s_{12} \,\,
 \frac{1}{2}) S \, M_{S} \, \rangle,  \label{eq.basis_spin}
\end{eqnarray}
and the isospin part $| \, \alpha_{T} \, \rangle$ is similar to the
spin part. As indicated in Fig.~\ref{fig.basis_states} the angular
dependence explicitly appears in the Jacobi vector variables,
whereas in a standard PW approach the angular dependence leads to
two orbital angular momentum quantum numbers, i.e. $l_{12}$ and
$l_{3}$ \cite{Stadler-PRL78}. It indicates that in the present 3D
formalism there is not any coupling between the orbital angular
momenta and the corresponding spin quantum numbers. Therefore we
couple the spin quantum numbers $s_{12}$ and $s_{3}$ to the total
spin $S$ and its third component $M_{S}$ as: $| \, (s_{12} \,\,
s_{3})S \, M_{S} \, \rangle$. For the isospin quantum numbers
similar coupling scheme leads to the total isospin $T, \, M_{T}$ as
$| \, (t_{12} \,\, t_{3})T \, M_{T} \, \rangle$.

In order to evaluate the transition and the permutation operators we
need the free 3N basis states $|\, {\bf p}\,{\bf q} \,\, \gamma \,
\rangle$, where
\begin{eqnarray}
 | \, \gamma \, \rangle \equiv| \, \gamma_{S} \, \gamma_{T} \,
 \rangle, \,\,\,\,\,
  | \, \gamma_{S} \, \rangle \equiv | \, m_{s_{1}} \, m_{s_{2}} \,
 m_{s_{3}} \, \rangle.
  \label{eq.basis_free}
\end{eqnarray}

The quantities $m_{s_{i}} \, (i=1,2,3)$ are the third components of
the spins of the three nucleons. The isospin part of the basis
states $| \, \gamma_{T} \, \rangle$ is similar to the spin part. To
achieve this aim when changing the 3N basis states $| \, \alpha \,
\rangle$ to the free 3N basis states $| \, \gamma \, \rangle$ we
need to calculate the following Clebsch-Gordan coefficients (see
appendix \ref{app:CG coeeficients}):
\begin{eqnarray}
\langle \, \gamma  | \, \alpha \, \rangle = g_{\gamma \alpha }
\equiv g_{\gamma \alpha }^{S} \, g_{\gamma \alpha }^{T} = \langle \,
m_{s_{1}} \, m_{s_{2}} \, m_{s_{3}}  | \, (s_{12} \,\, \frac{1}{2})
S \, M_{S} \, \rangle \, \langle \, m_{t_{1}} \, m_{t_{2}} \,
m_{t_{3}}  | \, (t_{12} \,\, \frac{1}{2}) T \, M_{T} \, \rangle.
  \label{eq.CG_coef}
\end{eqnarray}

The introduced basis states are complete and normalized as:
\begin{eqnarray}
\sum_{\xi} \int d^{3}p \, \int d^{3}q \,\,|\, {\bf p} \, {\bf q} \,
\xi \,\rangle \,\langle\, {\bf p} \, {\bf q} \,  \xi \,|=\mathbf{1}
 , \,\,\, \langle \, {\bf p} \, {\bf q} \,\xi\, |\,{\bf p}' \, {\bf
q}' \, \xi'\, \rangle=\delta^{3}({\bf p}-{\bf p}') \,
\delta^{3}({\bf q}-{\bf q}') \,\delta_{\xi\,\xi'},
\label{eq.normalization}
\end{eqnarray}
where $\xi$ indicates $\alpha$ and $\gamma$ quantum number sets. Now
we can represent the Eq. (\ref{eq.FC}) with respect to the basis
states which have been already introduced in  Eq. (\ref{eq.basis}):
\begin{eqnarray}
 \langle \, {\bf p}\,{\bf q}\, \alpha \,|\psi\rangle &=& \sum_{\alpha'} \, \int
d^{3}p' \, \int d^{3}q' \, \langle \, {\bf p}\,{\bf q} \, \alpha
\,|G_{0}t P| \, {\bf p}'\,{\bf q}' \, \alpha' \,\rangle \langle {\bf
p}'\,{\bf q}' \, \alpha'\, |\psi\rangle.
  \label{eq.FC_integral}
\end{eqnarray}

For evaluating the Eq. (\ref{eq.FC_integral}), we need to evaluate
the matrix elements of $\langle \, {\bf p}\,{\bf q}\, \alpha \,
|G_{0}t P | \, {\bf p}'\,{\bf q}' \, \alpha' \, \rangle$, towards
this aim, it is convenient to insert the free 3N completeness
relations as:
\begin{eqnarray}
  \langle \, {\bf p}\,{\bf q}\, \alpha \, |G_{0}t P | \, {\bf
p}'\,{\bf q}' \, \alpha' \, \rangle &=&
  \sum_{\gamma} \, \sum_{\gamma'} \, \langle \,
\alpha \, | \, \gamma \, \rangle \langle \, {\bf p}\,{\bf q}\,
\gamma \, |G_{0}t P | \, {\bf p}' \,{\bf q}' \, \gamma' \, \rangle
\langle \, \gamma' \, |  \, \alpha' \, \rangle \nonumber
\\* &=&
\sum_{\gamma,\gamma'}  \, g_{\alpha \gamma} \, g_{\gamma' \alpha'}
\langle \, {\bf p}\,{\bf q}\, \gamma \, |G_{0}t P | \, {\bf p}'
\,{\bf q}' \, \gamma' \, \rangle.
 \label{eq.g0tp}
\end{eqnarray}
For evaluating the matrix elements $\langle \, {\bf p}\,{\bf q}\,
\gamma \, |G_{0}t P |  \, {\bf p}' \,{\bf q}' \, \gamma' \, \rangle
$ we should insert again a free 3N completeness relation between the
between the two-nucleon $t$-matrix and the permutation operators as:
\begin{eqnarray}
 \langle \, {\bf p}\,{\bf q}\, \gamma \, |G_{0}t P | \,
{\bf p}'\,{\bf q}' \, \gamma' \, \rangle &=&
\frac{1}{E-\frac{p^{2}}{m} -\frac{3q^{2}}{4m}} \, \sum_{\gamma''}
\, \int d^{3}p'' \, \int d^{3}q''  \nonumber
\\* &\times&
\langle \, {\bf p}\,{\bf q} \, \gamma \, |t| \, {\bf p}''\,{\bf
q}'' \, \gamma'' \, \rangle
 \langle \, {\bf p}''\,{\bf q}'' \, \gamma'' \, | P| \, {\bf
p}'\,{\bf q}' \, \gamma' \, \rangle, \label{eq.g0tp_free}
\end{eqnarray}
where the matrix elements of the two-body $t$-matrix and the
permutation operator $P$ are evaluated separately as:
\begin{eqnarray}
\langle \, {\bf p}\,{\bf q} \, \gamma \, |t | \, {\bf p}''\,{\bf
q}'' \, \gamma'' \, \rangle &=& \delta^{3}({\bf q}-{\bf q}'')\,
\delta_{m_{s_{3}} m''_{s_{3}}} \, \delta_{m_{t_{3}} m''_{t_{3}}} \,
\langle \, {\bf p} \, m_{s_{1}} m_{s_{2}} \, m_{t_{1}} m_{t_{2}}
|t(\epsilon) |{\bf p}'' \, m''_{s_{1}} m''_{s_{2}} \, m''_{t_{1}}
m''_{t_{2}} \rangle, \nonumber
\\* \label{eq.t}
\end{eqnarray}
\begin{eqnarray}
 &&\langle \, {\bf p}''\,{\bf q}'' \, \gamma'' \,| P| \, {\bf
p}'\,{\bf q}' \, \gamma' \, \rangle  \nonumber
\\* &&=\delta^{3}({\bf p}''+\frac{1}{2}{\bf p}'+ \frac{3}{4}{\bf q}')
\, \delta^{3}({\bf q}''-{\bf p}'+ \frac{1}{2}{\bf q}')
 \, \delta_{m''_{s_{1}} m'_{s_{2}}} \, \delta_{m''_{s_{2}} m'_{s_{3}}} \, \delta_{m''_{s_{3}} m'_{s_{1}}}
 \, \delta_{m''_{t_{1}} m'_{t_{2}}} \, \delta_{m''_{t_{2}} m'_{t_{3}}} \, \delta_{m''_{t_{3}} m'_{t_{1}}}
 \nonumber \\* && +
  \delta^{3}({\bf p}''+\frac{1}{2}{\bf p}'-
\frac{3}{4}{\bf q}') \, \delta^{3}({\bf q}''+{\bf p}'+
\frac{1}{2}{\bf q}') \, \delta_{m''_{s_{1}} m'_{s_{3}}} \,
\delta_{m''_{s_{2}} m'_{s_{1}}} \, \delta_{m''_{s_{3}} m'_{s_{2}}}
\, \delta_{m''_{t_{1}} m'_{t_{3}}} \, \delta_{m''_{t_{2}}
m'_{t_{1}}} \, \delta_{m''_{t_{3}} m'_{t_{2}}},
  \nonumber \\* \label{eq.p}
\end{eqnarray}
where the two-body subsystem energy in the NN $t$-matrix is
$\epsilon = E-\frac{3q^{2}}{4m}$.

 In order to evaluate the
matrix elements of the permutation operator $P$ we have used the
relation between the Jacobi momenta in the different 3N systems
$(312),(231)$ and $(123)$. Inserting Eqs. (\ref{eq.t}) and
$(\ref{eq.p})$ into Eq. (\ref{eq.g0tp_free}) leads to:
\begin{eqnarray}
&& \langle \, {\bf p}\,{\bf q}\, \gamma \,  |G_{0}t P | \, {\bf
p}'\,{\bf q}' \, \gamma' \, \rangle = \frac{1}{E-\frac{p^{2}}{m}
-\frac{3q^{2}}{4m}} \nonumber
\\* &&  \times  \Biggl \{ \,\, \delta^{3}({\bf q}-{\bf p}'+ \frac{1}{2}{\bf q}')\,
\delta_{m_{s_{3}} m'_{s_{1}}} \, \delta_{m_{t_{3}} m'_{t_{1}}} \,
\langle {\bf p} \, m_{s_{1}} m_{s_{2}} \, m_{t_{1}} m_{t_{2}}
|t(\epsilon) |\frac{-1}{2}{\bf q}-{\bf q}' \, m'_{s_{2}}
m'_{s_{3}} \, m'_{t_{2}} m'_{t_{3}} \rangle \nonumber
\\* &&  \,\,\,\,\,\,\,\,\,\,+ \delta^{3}({\bf q}
+{\bf p}'+ \frac{1}{2}{\bf q}' )\, \delta_{m_{s_{3}} m'_{s_{2}}} \,
\delta_{m_{t_{3}} m'_{t_{2}}} \, \langle{\bf p}\, m_{s_{1}}
m_{s_{2}} \, m_{t_{1}} m_{t_{2}} |t(\epsilon) |\frac{1}{2}{\bf
q}+{\bf q}' \, m'_{s_{3}} m'_{s_{1}} \, m'_{t_{3}} m'_{t_{1}}
\rangle \,\, \Biggr\}. \nonumber
\\* \label{eq.g0tp-detail}
\end{eqnarray}

Inserting Eq. (\ref{eq.g0tp-detail}) into Eq. (\ref{eq.g0tp}) and
consequently inserting into Eq. (\ref{eq.FC_integral}) and
integrating over ${\bf p'}$ variable yields:
\begin{eqnarray}
&& \langle \, {\bf p}\,{\bf q}\, \alpha \, |\psi\rangle =
\frac{1}{{E-\frac{p^{2}}{m} -\frac{3q^{2}}{4m}}} \, \sum_{\gamma,
\gamma',\alpha'}  \,\, g_{\alpha \gamma} \, \, g_{\gamma' \alpha'}
\,  \int d^{3}q'
   \nonumber \\*   &&  \times \Biggl \{ \, \langle{\bf p}\, m_{s_{1}}
m_{s_{2}} \, m_{t_{1}} m_{t_{2}} |t(\epsilon) |\frac{-1}{2}{\bf
q}-{\bf q}' \, m'_{s_{2}} m'_{s_{3}} \, m'_{t_{2}} m'_{t_{3}}
\rangle  \,\,  \delta_{m_{s_{3}} m'_{s_{1}}} \, \delta_{m_{t_{3}}
m'_{t_{1}}}  \langle{\bf q}+\frac{1}{2}{\bf q}' \,\, {\bf q}' \,
\alpha'|\psi\rangle \nonumber \\* &&  \,\,\,\,\,\,\,\, + \langle{\bf
p}\, m_{s_{1}} m_{s_{2}} \, m_{t_{1}} m_{t_{2}} |t(\epsilon)
|\frac{1}{2}{\bf q}+{\bf q}' \, m'_{s_{3}} m'_{s_{1}} \, m'_{t_{3}}
m'_{t_{1}} \rangle \,\,  \delta_{m_{s_{3}} m'_{s_{2}}} \,
\delta_{m_{t_{3}} m'_{t_{2}}} \langle{\bf -q}-\frac{1}{2}{\bf q}'
\,\, {\bf q}' \, \alpha'|\psi\rangle \, \Biggl \}.  \nonumber \\*
 \label{eq.FC_p_applied}
 \end{eqnarray}

Applying the permutation operator $P_{12}$ action on the Faddeev
component, the space and also the spin-isospin parts of the basis
states, results in:
\begin{eqnarray}
P_{12} |\psi\rangle &=& -|\psi\rangle, \nonumber \\*P_{12} | \, {\bf
p}\,{\bf q} \rangle &=& | \, {\bf -p}\,{\bf q} \rangle, \nonumber
\\*P_{12} | \, \alpha \rangle &=&
(-)^{s_{1}+s_{2}-s_{12}}(-)^{t_{1}+t_{2}-t_{12}} | \, \alpha \rangle
= (-)^{s_{12}+t_{12}} |\, \alpha \rangle, \nonumber
\\* P_{12} | \,  \gamma \rangle  &=&  | \, m_{s_{2}} m_{s_{1}} m_{s_{3}} \,
m_{t_{2}} m_{t_{1}} m_{t_{3}}  \rangle,
  \label{eq.p12_application}
\end{eqnarray}
and consequently the following relations would be concluded:
\begin{eqnarray}
 \langle \, {\bf p}\,{\bf q}\, \alpha
\,|\psi\rangle &=& - (-)^{s_{12}+t_{12}} \langle -{\bf p}\,{\bf q}\,
\alpha \,|\psi\rangle, \nonumber \\*
 \langle{\bf p}\, m_{s_{1}} m_{s_{2}} \, m_{t_{1}} m_{t_{2}} |t(\epsilon) | {\bf
p}' \, m'_{s_{1}} m'_{s_{2}} \, m'_{t_{1}} m'_{t_{2}} \rangle &=&
\langle{\bf p}\, m_{s_{1}} m_{s_{2}} \, m_{t_{1}} m_{t_{2}}
|t(\epsilon) \, P_{12}| {\bf -p}' \, m'_{s_{2}} m'_{s_{1}} \,
m'_{t_{2}} m'_{t_{1}} \rangle.  \nonumber \\*
  \label{eq.p12_relations}
\end{eqnarray}

Therefore, we can rewrite Eq. (\ref{eq.FC_p_applied}) as:
\begin{eqnarray}
 \langle \, {\bf p}\,{\bf q}\, \alpha \, |\psi\rangle &=&
\frac{1}{{E-\frac{p^{2}}{m} -\frac{3q^{2}}{4m}}} \, \sum_{\gamma,
\gamma',\alpha'}  \,\, g_{\alpha \gamma} \, \, g_{\gamma' \alpha'}
\,  \int d^{3}q'
   \nonumber \\*   &\times&   \Biggl \{ \, \langle{\bf p}\, m_{s_{1}}
m_{s_{2}} \, m_{t_{1}} m_{t_{2}} |t(\epsilon) |\frac{-1}{2}{\bf
q}-{\bf q}' \, m'_{s_{2}} m'_{s_{3}} \, m'_{t_{2}} m'_{t_{3}}
\rangle   \,\, \delta_{m_{s_{3}} m'_{s_{1}}} \, \delta_{m_{t_{3}}
m'_{t_{1}}} \nonumber \\*   && \quad \times   \langle{\bf
q}+\frac{1}{2}{\bf q}' \,\, {\bf q}' \, \alpha'|\psi\rangle
\nonumber \\* && \,\, + \langle{\bf p}\, m_{s_{1}} m_{s_{2}} \,
m_{t_{1}} m_{t_{2}} |t(\epsilon) \, P_{12} |\frac{-1}{2}{\bf q}-{\bf
q}' \, m'_{s_{1}} m'_{s_{3}} \, m'_{t_{1}} m'_{t_{3}} \rangle \,\,
\delta_{m_{s_{3}} m'_{s_{2}}} \, \delta_{m_{t_{3}} m'_{t_{2}}}
\nonumber \\*   && \quad \times  \biggl (- (-)^{s_{12}'+t_{12}'}
\biggr) \langle{\bf q}+\frac{1}{2}{\bf q}' \,\, {\bf q}' \,
\alpha'|\psi\rangle \, \Biggl \}  \nonumber \\* &=&
\frac{1}{{E-\frac{p^{2}}{m} -\frac{3q^{2}}{4m}}} \, \sum_{\gamma,
\gamma',\alpha'} \,  g_{\alpha \gamma} \, \, g_{\gamma' \alpha'} \,
 \delta_{m_{s_{3}} m'_{s_{1}}} \, \delta_{m_{t_{3}}
m'_{t_{1}}}  \int d^{3}q' \nonumber
\\*   &\times&    \langle{\bf p}\, m_{s_{1}}
m_{s_{2}} \, m_{t_{1}} m_{t_{2}} |t(\epsilon) (1-P_{12})
|\frac{-1}{2}{\bf q}-{\bf q}' \, m'_{s_{2}} m'_{s_{3}} \, m'_{t_{2}}
m'_{t_{3}} \rangle   \,\,  \langle{\bf q}+\frac{1}{2}{\bf q}' \,\,
{\bf q}' \, \alpha'|\psi\rangle. \nonumber \\*
 \label{eq.FC_p12_applied}
 \end{eqnarray}

The final derivation of Eq. (\ref{eq.FC_p12_applied}) is made by the
exchange of labels $m'_{s_{1}}, m'_{t_{1}}$ to $m'_{s_{2}},
m'_{t_{2}}$ and reverse of it in the second term as well as the
following relation;
\begin{eqnarray}
 g_{\gamma' \alpha'}= (-)^{s_{12}'+t_{12}'} \, \langle \, m_{s_{2}}' \, m_{s_{1}}' \,
m_{s_{3}}'  | \, (s_{12}' \,\, \frac{1}{2}) S' \, M_{S}' \, \rangle
\, \langle \, m_{t_{2}}' \, m_{t_{1}}' \, m_{t_{3}}'  | \, (t_{12}'
\,\, \frac{1}{2}) T' \, M_{T}' \, \rangle. \label{eq.g_alpha_gamma_}
\end{eqnarray}

By introducing the physical representation of the two-body
$t$-matrix follows (see appendix \ref{app:t matrix});
\begin{eqnarray}
  _{a}\langle{\bf p} \, m_{s_{1}} m_{s_{2}} \, m_{t_{1}}
m_{t_{2}} |t(\varepsilon)| {\bf p}' \, m'_{s_{1}} m'_{s_{2}} \,
m'_{t_{1}} m'_{t_{2}} \rangle _{a} && \nonumber
\\* && \hspace{-30mm}= \langle{\bf p} \, m_{s_{1}} m_{s_{2}}
\, m_{t_{1}} m_{t_{2}} |t(\varepsilon)(1-P_{12})| {\bf p}' \,
m'_{s_{1}} m'_{s_{2}} \, m'_{t_{1}} m'_{t_{2}} \rangle,
\label{eq.ta}
\end{eqnarray}
the three-dimensional Faddeev integral equations can be obtained as:
\begin{eqnarray}
&& \langle \, {\bf p}\,{\bf q}\, \alpha \, |\psi\rangle =
\frac{1}{{E-\frac{p^{2}}{m} -\frac{3q^{2}}{4m}}} \, \sum_{\gamma,
\gamma',\alpha'} \,  g_{\alpha \gamma} \, \, g_{\gamma' \alpha'}
\,
 \delta_{m_{s_{3}} m'_{s_{1}}} \, \delta_{m_{t_{3}}
m'_{t_{1}}}   \nonumber
\\*   && \,\,\,\,\,\,\,\, \times \int d^{3}q' \,\, _{a}\langle{\bf p}\, m_{s_{1}}
m_{s_{2}} \, m_{t_{1}} m_{t_{2}} |t(\epsilon) |\frac{-1}{2}{\bf
q}-{\bf q}' \, m'_{s_{2}} m'_{s_{3}} \, m'_{t_{2}} m'_{t_{3}}
\rangle_{a}   \,\,  \langle{\bf q}+\frac{1}{2}{\bf q}' \,\, {\bf q}'
\, \alpha'|\psi\rangle. \,\,\,\,
 \label{eq.FC-final}
 \end{eqnarray}

The Faddeev component $\langle \, {\bf p}\,{\bf q}\, \alpha \,
|\psi\rangle$ is given as a function of Jacobi momenta vectors,
${\bf p}$ and ${\bf q}$, and also quantum number sets, $\alpha$, as
a solution of the spatial three-dimensional integral equations, Eq.
(\ref{eq.FC-final}). In order to solve this equation directly and
without employing the PW projections, we have to define a coordinate
system. It is convenient to choose the spin polarization direction
parallel to the $z$-axis and express the momentum vectors in this
coordinate system. By these considerations we can rewrite Eq.
(\ref{eq.FC-final}) as:
\begin{eqnarray}
\psi^{\alpha}(p\,\,q\,\,x_{pq}) &=& \frac{1}{{E-\frac{p^{2}}{m}
-\frac{3q^{2}}{4m}}} \,  \int_{0}^{\infty} dq' \, q'^{2}
\int_{-1}^{+1} dx_{q'} \int_{0}^{2\pi} d\varphi_{q'} \nonumber
\\* &\times&  \,
\sum_{\alpha'} \, T_{\alpha \alpha'}(p,\tilde{\pi}, x_{p
\tilde{\pi}} ;\epsilon) \, \psi^{\alpha'}(\pi\,\, q'\,\, x_{\pi
q'}),
 \label{eq.FC-magnitude}
 \end{eqnarray}
where
\begin{eqnarray}
T_{\alpha \alpha'}(p,\tilde{\pi}, x_{p \tilde{\pi}}
;\epsilon)=\sum_{\gamma,\gamma'} \, g_{\alpha \gamma} \, g_{\gamma'
\alpha'} \, \delta_{m_{s_{3}} m'_{s_{1}}} \, \delta_{m_{t_{3}}
m'_{t_{1}}} \,   t_{a}\, _{\, m_{s_{1}} m_{s_{2}} \, m_{t_{1}}
m_{t_{2}}} ^{\, m'_{s_{2}} m'_{s_{3}} \, m'_{t_{2}}
m'_{t_{3}}}(p,\tilde{\pi}, x_{p \tilde{\pi}} ;\epsilon),
 \label{eq.kernel}
 \end{eqnarray}
\begin{eqnarray}
x_{pq}&=& x_{p}x_{q}+\sqrt{1-x_{p}^{2}}\sqrt{1-x_{q}^{2}}\sin
(\phi_{p}-\phi_{q}), \nonumber
\\*
x_{pq'}&=& x_{p}x_{q'}+\sqrt{1-x_{p}^{2}}\sqrt{1-x_{q'}^{2}}\sin
(\phi_{p}-\phi_{q'}), \nonumber
\\*
x_{qq'}&=& x_{q}x_{q'}+\sqrt{1-x_{q}^{2}}\sqrt{1-x_{q'}^{2}}\sin
(\phi_{q}-\phi_{q'}), \nonumber
\\*
\tilde{\pi}&=&\sqrt{\frac{1}{4}q^{2}+q'^{2}+qq'x_{qq'}}, \nonumber
\\*
 x_{p \tilde{\pi}}
&=&\frac{\frac{1}{2}qx_{pq}+q'x_{pq'}}{\tilde{\pi}}, \nonumber
\\* \pi&=&\sqrt{q^{2}+\frac{1}{4}q'^{2}+qq'x_{qq'}}, \nonumber
\\* x_{\pi q'}&=&\frac{qx_{qq'}+\frac{1}{2}q'}{\pi}. \label{eq.variables}
 \end{eqnarray}

 In a standard PW approach, Eq.~(\ref{eq.FC-magnitude}) is replaced by
a set of an infinite number of coupled two-dimensional integral
equations for the amplitudes with the kernels containing relatively
complicated geometrical expressions:
\begin{eqnarray}
\psi^{\alpha}(p\,\,q) &=& \frac{1}{{E-\frac{p^{2}}{m}
-\frac{3q^{2}}{4m}}} \int_{0}^{\infty} dq' \, q'^{2}
\int_{-1}^{+1} dx_{q'} \nonumber
\\* \, &\times& \sum_{l_{12}'',\alpha'} \, \frac{t_{l_{12}l''_{12}}^{s_{12}j_{12}t_{12}}(p,\tilde{\pi}
;\epsilon)}{\tilde{\pi}^{l_{12}''}}   \,
G_{\alpha\alpha'}(q,q',x_{q'}) \, \frac{\psi^{\alpha'}(\pi\,\,
q')}{\pi^{l_{12}''}}, \label{eq.FC-PW}
 \end{eqnarray}
where, as is shown in Fig.~\ref{fig.basis_states}, the spin-space as
well as the isospin parts of the basis states in the PW
decomposition are $| \alpha \rangle \equiv |((l_{12}s_{12})j_{12}\,
(l_{3}s_{3})j_{3})JM_{J} \, (t_{12}t_{3})TM_{T}\rangle$.
$G_{\alpha\alpha'}(q,q',x_{q'})$ is composed of Legendre functions,
powers of $q$ and $q'$ and purely complicated geometrical quantities
like Clebsch-Gordan coefficients and $6j$ symbols. The comparison of
Eqs. (\ref{eq.FC-magnitude}) and (\ref{eq.FC-PW}) shows that new 3D
formalism avoids the highly involved angular momentum algebra
occurring for the permutations and additionally it will be more
efficient especially for the three-body forces
\cite{Hadizadeh-preparation}.

\subsection{The 3N Wave Function} \label{subsec:3N wave function}
The representation of the total wave function, Eq. (\ref{eq.WF}),
with respect to the basis states which have been introduced in Eq.
(\ref{eq.basis}), reads as follows:
\begin{eqnarray}
\langle  \, {\bf p} \, {\bf q} \,\,  \alpha  \, |\Psi \rangle &=&
\langle  \, {\bf p} \, {\bf q} \,\,  \alpha  \, |(1+P) |\psi \rangle
\nonumber \\*  &=& \langle  \, {\bf p} \, {\bf q} \,\, \alpha \,
|\psi \rangle   + \langle  \, {\bf p} \, {\bf q} \,\, \alpha \,
|P_{12}P_{23}|\psi \rangle + \langle  \, {\bf p} \, {\bf q} \,\,
\alpha \, |P_{13}P_{23}|\psi \rangle, \label{eq.WF-operator}
\end{eqnarray}
where the first Faddeev component
\begin{eqnarray}
 \langle  \, {\bf p} \, {\bf q} \,\, \alpha \, |\psi \rangle
 \equiv  \,\, _{3}\langle  \, {\bf p} \, {\bf q} \,\, \alpha \,
 |\psi \rangle &\equiv&  \psi^{\alpha} ({\bf p} \,, {\bf q})
 \equiv \psi^{\alpha} (p\,\,q\,\,x_{pq}), \label{eq.1th-FC}
\end{eqnarray}
is given explicitly as a three-dimensional integral equation, Eq.
(\ref{eq.FC-magnitude}). Here the subscript $3$ of the bra basis
states stands for the three-body subsystem $(12,3)$, which as matter
of convenience, is called subsystem $3$. For the second and third
components we need to evaluate the action of the cyclic and the
anti-cyclic permutation operators $P_{12}P_{23}$ and $P_{13}P_{23}$
on the first component as:
\begin{eqnarray}
 \langle  \, {\bf p} \, {\bf q} \,\, \alpha \, |P_{12}P_{23} |\psi
 \rangle &\equiv&  \,\, _{3}\langle \, {\bf p} \, {\bf q} \,\,
 \alpha \, |P_{12}P_{23}|\psi \rangle \nonumber \\* &=&
 \sum_{\alpha'} \, \int d^{3}p' \, \int d^{3}q' \, _{3}\langle \,
 {\bf p}\,{\bf q} \, \alpha \,| P_{12}P_{23} | \, {\bf p}'\,{\bf
 q}' \, \alpha' \,\rangle_{3} \,\, _{3} \langle {\bf p}'\,{\bf q}'
 \, \alpha'\, |\psi\rangle \nonumber \\* &=& \sum_{\alpha'} \,
 \int d^{3}p' \, \int d^{3}q' \, _{3}\langle \, {\bf p}\,{\bf q}
 \, \alpha \, | \, {\bf p}'\,{\bf q}' \, \alpha' \,\rangle_{1}
 \,\, _{3} \langle {\bf p}'\,{\bf q}' \, \alpha'\, |\psi\rangle,
 \nonumber \\*\nonumber \\* \langle  \, {\bf p} \, {\bf q} \,\,
 \alpha \, |P_{13}P_{23} |\psi \rangle &\equiv&  \,\, _{3}\langle
 \, {\bf p} \, {\bf q} \,\, \alpha \, |P_{13}P_{23}|\psi \rangle
 \nonumber \\* &=&  \sum_{\alpha'} \, \int d^{3}p' \, \int d^{3}q'
 \, _{3}\langle \, {\bf p}\,{\bf q} \, \alpha \,| P_{13}P_{23} |
 \, {\bf p}'\,{\bf q}' \, \alpha' \,\rangle_{3} \,\, _{3} \langle
 {\bf p}'\,{\bf q}' \, \alpha'\, |\psi\rangle \nonumber \\* &=&
 \sum_{\alpha'} \, \int d^{3}p' \, \int d^{3}q' \, _{3}\langle \,
 {\bf p}\,{\bf q} \, \alpha \, | \, {\bf p}'\,{\bf q}' \, \alpha'
 \,\rangle _{2} \,\, _{3} \langle {\bf p}'\,{\bf q}' \, \alpha'\,
 |\psi\rangle, \label{eq.P12P23-P13P23}
\end{eqnarray}
the space as well as the spin-isospin parts of the coordinate
transformations $_{3}\langle  \, |  \, \rangle_{1} $ and
$_{3}\langle  \, |  \, \rangle_{2} $ can be evaluated as:
\begin{eqnarray}
 _{3}\langle \, {\bf p}\,{\bf q} \, \alpha \, | \, {\bf p}'\,{\bf
 q}' \, \alpha' \,\rangle_{1} &=& _{3}\langle \, {\bf p}\,{\bf q}
 \, | \, {\bf p}'\,{\bf q}' \,\rangle_{1} \,\, _{3}\langle \,
 \alpha \, | \, \alpha' \,\rangle_{1} \nonumber
\\*
 &=& \delta^{3}({\bf p}'+\frac{1}{2}{\bf p}+ \frac{3}{4}{\bf q})
 \, \delta^{3}({\bf q}'-{\bf p}+ \frac{1}{2}{\bf q}) \nonumber \\*
 &\times&\delta_{M_{S}M_{S}'} \, \delta_{SS'} \,
 \delta_{M_{T}M_{T}'} \, \delta_{TT'} \,
 C^{*}_{S}(\alpha_{S},s'_{23}) \,  C^{*}_{T}(\alpha_{T},t'_{23}),
 \label{eq.transformations31}
\end{eqnarray}
\begin{eqnarray}
 _{3}\langle \, {\bf p}\,{\bf q} \, \alpha \, | \, {\bf p}'\,{\bf
 q}' \, \alpha' \,\rangle_{2} &=& \, _{3}\langle \, {\bf p}\,{\bf
 q} \, | \, {\bf p}'\,{\bf q}' \,\rangle_{2} \,\, _{3}\langle \,
 \alpha \, | \, \alpha' \,\rangle_{2} \nonumber \\* &=&
 \delta^{3}({\bf p}'+\frac{1}{2}{\bf p}- \frac{3}{4}{\bf q}) \,
 \delta^{3}({\bf q}'+{\bf p}+ \frac{1}{2}{\bf q}) \nonumber \\*
 &\times&  \,\delta_{M_{S}M_{S}'} \, \delta_{SS'} \,
 \delta_{M_{T}M_{T}'} \, \delta_{TT'} \,
 C^{**}_{S}(\alpha_{S},s'_{31}) \, C^{**}_{T}(\alpha_{T},t'_{31}),
 \label{eq.transformations32}
\end{eqnarray}
where the spin coefficients $C^{*}_{S}$ and $C^{**}_{S}$ are given
as:
\begin{eqnarray}
 C^{*}_{S}(\alpha_{S},s'_{23}) &=& (-)^{s_{23}'+2s_{1}+s_{2}+s_{3}} \, \left\{%
\begin{array}{ccc}
  s_{1} & s_{2} & s_{12} \\
  s_{3} & S & s_{23}' \\
\end{array}%
\right\}, \nonumber \\*
C^{**}_{S}(\alpha_{S},s'_{31}) &=& (-)^{s_{31}'+2s_{2}+s_{3}+s_{1}} \, \left\{%
\begin{array}{ccc}
  s_{1} & s_{2} & s_{12} \\
  s_{3} & S & s_{31}' \\
\end{array}%
\right\},  \label{eq.spin-coeefs}
\end{eqnarray}
and the isospin coefficients $C^{*}_{T}$ and $C^{**}_{T}$ are
similar to the corresponding spin coefficients. By these
considerations we obtain the second and third Faddeev components as:
\begin{eqnarray}
 \langle  \, {\bf p} \, {\bf q} \,\, \alpha \, |P_{12}P_{23} |\psi
 \rangle &=&  \sum_{s'_{23},t'_{23}} \,
 C^{*}_{S}(\alpha_{S},s'_{23}) \,  C^{*}_{T}(\alpha_{T},t'_{23})
 \, \psi^{\alpha^{*}} (-\frac{1}{2}{\bf p}- \frac{3}{4}{\bf q} \,,
 {\bf p}- \frac{1}{2}{\bf q}) \nonumber \\* &\equiv&
 \sum_{s'_{23},t'_{23}} \, C^{*}_{S}(\alpha_{S},s'_{23}) \,
 C^{*}_{T}(\alpha_{T},t'_{23}) \, \psi^{\alpha^{*}} (\pi_{1} \,\,
 \pi_{2} \,\, x_{\pi_{1} \pi_{2}}),  \nonumber
 \\* \nonumber
 \\* \langle  \, {\bf p} \, {\bf q} \,\, \alpha \, |P_{13}P_{23}
 |\psi \rangle &=&  \sum_{s'_{31},t'_{31}} \,
 C^{**}_{S}(\alpha_{S},s'_{31}) \, C^{**}_{T}(\alpha_{T},t'_{31})
 \, \psi^{\alpha^{**}} (-\frac{1}{2}{\bf p}+\frac{3}{4}{\bf q} \,,
 -{\bf p}- \frac{1}{2}{\bf q})  \nonumber \\* &\equiv&
 \sum_{s'_{31},t'_{31}} \, C^{**}_{S}(\alpha_{S},s'_{31}) \,
 C^{**}_{T}(\alpha_{T},t'_{31}) \, \psi^{\alpha^{**}} (\Pi_{1}
 \,\, \Pi_{2} \,\, x_{\Pi_{1}\Pi_{2}} ),  \label{eq.2th-3th-FCs}
\end{eqnarray}
where
\begin{eqnarray}
 | \, \alpha ^{*}\, \rangle &=& | \, (s'_{23} \,\, \frac{1}{2}) S
 \, M_{S} \, \, (t'_{23} \,\, \frac{1}{2}) T \, M_{T} \, \rangle,
 \nonumber \\* \pi_{1}&=&
 \sqrt{\frac{1}{4}p^{2}+\frac{9}{16}q^{2}+\frac{3}{4}p \, q \,
 x_{pq} }, \nonumber \\* \pi_{2}&=& \sqrt{p^{2}+\frac{1}{4}q^{2}-p
 \, q \, x_{pq} }, \nonumber \\* x_{\pi_{1} \pi_{2}}&=& \frac{1}{
 \pi_{1}\pi_{2}} (-\frac{1}{2} p^{2} + \frac{3}{8} q^{2}
 -\frac{1}{2} p \, q\, x_{pq}),  \label{eq.2th-FCs-variables}
\end{eqnarray}

\begin{eqnarray}
  | \, \alpha^{**}\, \rangle &=& | \, (s'_{31} \,\, \frac{1}{2}) S
 \, M_{S} \, \, (t'_{31} \,\, \frac{1}{2}) T \, M_{T} \, \rangle,
 \nonumber \\* \Pi_{1}&=&
 \sqrt{\frac{1}{4}p^{2}+\frac{9}{16}q^{2}-\frac{3}{4}p \, q \,
 x_{pq} }, \nonumber \\* \Pi_{2}&=& \sqrt{p^{2}+\frac{1}{4}q^{2}+p
 \, q \, x_{pq} }, \nonumber \\* x_{\Pi_{1} \Pi_{2}}&=& \frac{1}{
 \Pi_{1}\Pi_{2}} (\frac{1}{2} p^{2} - \frac{3}{8} q^{2} -
 \frac{1}{2} p \, q\, x_{pq}). \label{eq.3th-FCs-variables}
\end{eqnarray}

\subsection{Comparison of Coupled Faddeev Equations in both 3D and PW Schemes } \label{subsec:discussion}
In this section we discuss the number of coupled equations in both
3D and PW approaches. In a standard PW approach the infinite set of
coupled integral equations, given in Eq. (\ref{eq.FC-PW}), is
truncated in the actual calculations at sufficiently high values of
the angular momentum quantum numbers. If one assumes that the NN
$t$-matrix acts only in very few partial waves then the number of
the coupled equations are correspondingly small. As shown in Table
\ref{table:No of PWs}, if NN $t$-matrix acts up to $j_{12}^{max}=1,
2, 3, 4$ and $5$, then the number of channels will be 5, 18, 26, 34
and 42. This is while the total isospin is restricted to
$T=\frac{1}{2}$ \cite{Machleidt-ANP19}.

\begin{table}[hbt]
\caption {The number of PW channels which compose the Triton wave
function when the NN $t$-matrix acts up to different total
two-nucleon angular momenta $j_{12}^{max}$. Total isospin is
restricted to $T=\frac{1}{2}$. The number of channels for
$j_{12}^{max}=1$, namely $N_{\alpha}=5$, is related to only positive
parity states.}
\begin{tabular}{ccccccccccccccccccccccccc}
\hline \hline
  $j_{12}^{max}$ &&&& 1 &&&& 2 &&&& 3 &&&& 4 &&&& 5  \\
\hline $N_{\alpha}$ &&&& 5 &&&& 18 &&&& 26 &&&& 34 &&&& 42 \\ \hline
\hline
\end{tabular}
\label{table:No of PWs}
\end{table}

\begin{table}[hbt]
\caption {Quantum numbers of the spin-isospin states which compose
$^{3}H$ or $^{3}He$ wave function. }
\begin{tabular}{ccccccccccc}
\hline \hline  channel && $(s_{12} \,\, \frac{1}{2}) S \,\, M_{S}$ && $(t_{12} \,\, \frac{1}{2}) T \,\, M_{T}$ && $(S-T)$ \\
\hline 1 && $(0 \,\, \frac{1}{2}) \frac{1}{2} \,\, \frac{+1}{2}$ &&
$(0 \,\, \frac{1}{2}) \frac{1}{2} \,\, \frac{+1}{2}/\frac{-1}{2}$ &&
$(\frac{1}{2}-\frac{1}{2})$  \\
2 && $(0 \,\, \frac{1}{2}) \frac{1}{2} \,\, \frac{-1}{2}$ && $(0
\,\, \frac{1}{2}) \frac{1}{2} \,\, \frac{+1}{2}/\frac{-1}{2}$ &&
$(\frac{1}{2}-\frac{1}{2})$  \\
3 && $(1 \,\, \frac{1}{2}) \frac{1}{2} \,\, \frac{+1}{2}$ && $(0
\,\, \frac{1}{2}) \frac{1}{2} \,\, \frac{+1}{2}/\frac{-1}{2}$ &&
$(\frac{1}{2}-\frac{1}{2})$  \\
4 && $(1 \,\, \frac{1}{2}) \frac{1}{2} \,\, \frac{-1}{2}$ && $(0
\,\, \frac{1}{2}) \frac{1}{2} \,\, \frac{+1}{2}/\frac{-1}{2}$ &&
$(\frac{1}{2}-\frac{1}{2})$  \\
5 && $(0 \,\, \frac{1}{2}) \frac{1}{2} \,\, \frac{+1}{2}$ && $(1
\,\, \frac{1}{2}) \frac{1}{2} \,\, \frac{+1}{2}/\frac{-1}{2}$ &&
$(\frac{1}{2}-\frac{1}{2})$  \\
6 && $(0 \,\, \frac{1}{2}) \frac{1}{2} \,\, \frac{-1}{2}$ && $(1
\,\, \frac{1}{2}) \frac{1}{2} \,\, \frac{+1}{2}/\frac{-1}{2}$ &&
$(\frac{1}{2}-\frac{1}{2})$  \\
7 && $(1 \,\, \frac{1}{2}) \frac{1}{2} \,\, \frac{+1}{2}$ && $(1
\,\, \frac{1}{2}) \frac{1}{2} \,\, \frac{+1}{2}/\frac{-1}{2}$ &&
$(\frac{1}{2}-\frac{1}{2})$  \\
8 && $(1 \,\, \frac{1}{2}) \frac{1}{2} \,\, \frac{-1}{2}$ && $(1
\,\, \frac{1}{2}) \frac{1}{2} \,\, \frac{+1}{2}/\frac{-1}{2}$ &&
$(\frac{1}{2}-\frac{1}{2})$  \\
\hline 9 && $(0 \,\, \frac{1}{2}) \frac{1}{2} \,\, \frac{+1}{2}$ &&
$(1 \,\, \frac{1}{2}) \frac{3}{2} \,\, \frac{+1}{2}/\frac{-1}{2}$ &&
$(\frac{1}{2}-\frac{3}{2})$  \\
10 && $(0 \,\, \frac{1}{2}) \frac{1}{2} \,\, \frac{-1}{2}$ && $(1
\,\, \frac{1}{2}) \frac{3}{2} \,\, \frac{+1}{2}/\frac{-1}{2}$ &&
$(\frac{1}{2}-\frac{3}{2})$  \\
11 && $(1 \,\, \frac{1}{2}) \frac{1}{2} \,\, \frac{+1}{2}$ && $(1
\,\, \frac{1}{2}) \frac{3}{2} \,\, \frac{+1}{2}/\frac{-1}{2}$ &&
$(\frac{1}{2}-\frac{3}{2})$  \\
12 && $(1 \,\, \frac{1}{2}) \frac{1}{2} \,\, \frac{-1}{2}$ && $(1
\,\, \frac{1}{2}) \frac{3}{2} \,\, \frac{+1}{2}/\frac{-1}{2}$ &&
$(\frac{1}{2}-\frac{3}{2})$  \\
\hline 13 && $(1 \,\, \frac{1}{2}) \frac{3}{2} \,\, \frac{+3}{2}$ &&
$(0 \,\, \frac{1}{2}) \frac{1}{2} \,\, \frac{+1}{2}/\frac{-1}{2}$ &&
$(\frac{3}{2}-\frac{1}{2})$  \\
14 && $(1 \,\, \frac{1}{2}) \frac{3}{2} \,\, \frac{+1}{2}$ && $(0
\,\, \frac{1}{2}) \frac{1}{2} \,\, \frac{+1}{2}/\frac{-1}{2}$ &&
$(\frac{3}{2}-\frac{1}{2})$  \\
15 && $(1 \,\, \frac{1}{2}) \frac{3}{2} \,\, \frac{-1}{2}$ && $(0
\,\, \frac{1}{2}) \frac{1}{2} \,\, \frac{+1}{2}/\frac{-1}{2}$ &&
$(\frac{3}{2}-\frac{1}{2})$  \\
16 && $(1 \,\, \frac{1}{2}) \frac{3}{2} \,\, \frac{-3}{2}$ && $(0
\,\, \frac{1}{2}) \frac{1}{2} \,\, \frac{+1}{2}/\frac{-1}{2}$ &&
$(\frac{3}{2}-\frac{1}{2})$  \\
17 && $(1 \,\, \frac{1}{2}) \frac{3}{2} \,\, \frac{+3}{2}$ && $(1
\,\, \frac{1}{2}) \frac{1}{2} \,\, \frac{+1}{2}/\frac{-1}{2}$ &&
$(\frac{3}{2}-\frac{1}{2})$  \\
18 && $(1 \,\, \frac{1}{2}) \frac{3}{2} \,\, \frac{+1}{2}$ && $(1
\,\, \frac{1}{2}) \frac{1}{2} \,\, \frac{+1}{2}/\frac{-1}{2}$ &&
$(\frac{3}{2}-\frac{1}{2})$  \\
19 && $(1 \,\, \frac{1}{2}) \frac{3}{2} \,\, \frac{-1}{2}$ && $(1
\,\, \frac{1}{2}) \frac{1}{2} \,\, \frac{+1}{2}/\frac{-1}{2}$ &&
$(\frac{3}{2}-\frac{1}{2})$  \\
20 && $(1 \,\, \frac{1}{2}) \frac{3}{2} \,\, \frac{-3}{2}$ && $(1
\,\, \frac{1}{2}) \frac{1}{2} \,\, \frac{+1}{2}/\frac{-1}{2}$ &&
$(\frac{3}{2}-\frac{1}{2})$  \\ \hline 21 && $(1 \,\, \frac{1}{2})
\frac{3}{2} \,\, \frac{+3}{2}$ && $(1 \,\, \frac{1}{2}) \frac{3}{2}
\,\, \frac{+1}{2}/\frac{-1}{2}$ &&
$(\frac{3}{2}-\frac{3}{2})$  \\
22 && $(1 \,\, \frac{1}{2}) \frac{3}{2} \,\, \frac{+1}{2}$ && $(1
\,\, \frac{1}{2}) \frac{3}{2} \,\, \frac{+1}{2}/\frac{-1}{2}$ &&
$(\frac{3}{2}-\frac{3}{2})$  \\
23 && $(1 \,\, \frac{1}{2}) \frac{3}{2} \,\, \frac{-1}{2}$ && $(1
\,\, \frac{1}{2}) \frac{3}{2} \,\, \frac{+1}{2}/\frac{-1}{2}$ &&
$(\frac{3}{2}-\frac{3}{2})$  \\
24 && $(1 \,\, \frac{1}{2}) \frac{3}{2} \,\, \frac{-3}{2}$ && $(1
\,\, \frac{1}{2}) \frac{3}{2} \,\, \frac{+1}{2}/\frac{-1}{2}$ &&
$(\frac{3}{2}-\frac{3}{2})$  \\
 \hline\hline
\end{tabular}
 \label{table_Faddeev_channels}
\end{table}

\begin{table}[hbt]
\caption {The number of spin-isospin states for 3N bound sates, i.e.
$^{3}H$ and $^{3}He$, in a realistic 3D formalism. $N_{S}$ and
$N_{T}$ are the number of spin and isospin states correspondingly.}
\begin{tabular}{cccccccc|cccccccccccc}
\hline \hline $(s_{12} \,\, \frac{1}{2}) S \,\, M_{S}$ &&
$S=\frac{1}{2}$ && $S= \frac{3}{2}$ && $S=\frac{1}{2}, \frac{3}{2}$
&&
$(t_{12} \,\, \frac{1}{2}) T \,\, M_{T}$ && $T=\frac{1}{2}$ && $T= \frac{3}{2}$ && $T=\frac{1}{2}, \frac{3}{2}$ \\
\hline $(0 \,\, \frac{1}{2})\frac{1}{2} \,\, \frac{\pm1}{2} $ && 2
&& 0 && 2+0 &&
$(0 \,\, \frac{1}{2})\frac{1}{2}  \,\, \frac{+1}{2}/\frac{-1}{2} $ && 1 && 0 && 1+0 \\
$(1 \,\, \frac{1}{2})\frac{1}{2}  \,\, \frac{\pm1}{2}$ && 2 && 0 &&
2+0 &&
$(1 \,\, \frac{1}{2})\frac{1}{2}  \,\,  \frac{+1}{2}/\frac{-1}{2} $ && 1 && 0 &&  1+0 \\
$(1 \,\, \frac{1}{2})\frac{3}{2}  \,\,  \frac{\pm1}{2}
\frac{\pm3}{2}$ && 0 && 4 && 0+4 &&
$(1 \,\, \frac{1}{2})\frac{3}{2} \,\, \frac{+1}{2}/\frac{-1}{2}$ && 0 && 1 && 0+1 \\
$N_{S}$ && 4 && 4 && 8 && $N_{T}$ && 2 && 1 && 3 \\
\hline\hline
\end{tabular}
\label{table_3D_states}
\end{table}

\begin{table}[hbt]
\caption {The number of coupled Faddeev equations for the 3N bound
state, i.e. $^{3}H$ and $^{3}He$, in a realistic 3D formalism
according to the spin-isospin states $(S-T)$. $N=N_{S} \times N_{T}$
is the total number of coupled Faddeev equations. The star
superscript indicates all the spin or isospin states that one can
take into account up to a specific value. }
\begin{tabular}{c|cccccccccc}
\hline \hline $(S-T)$ && $(\frac{1}{2}-\frac{1}{2})$ &&
$(\frac{1}{2}-\frac{3}{2}^{\ast})$  &&
$(\frac{3}{2}^{\ast}-\frac{1}{2})$ &&
$(\frac{3}{2}^{\ast}-\frac{3}{2}^{\ast})$ \\
\hline $N_{S}$ &&
4 &&  4 && 8 && 8  \\
$N_{T}$ &&
2 &&  3 && 2 && 3  \\
$N$  &&
8 &&  12 && 16 && 24  \\
\hline\hline
\end{tabular}
\label{table_3D_Ystates}
\end{table}

In Table \ref{table_Faddeev_channels} we list all the spin-isospin
states which compose the 3N, i.e. $^{3}H$ and $^{3}He$, wave
function and consequently in Tables \ref{table_3D_states} and
\ref{table_3D_Ystates} we present the number of spin-isospin states
for the 3N bound states as well as the number of coupled Faddeev
equations in realistic 3D formalism presented in this paper. It is
clear that $M_{T}=\frac{+1}{2}$ refers to $^{3}He$ and
$M_{T}=\frac{-1}{2}$ refers to $^{3}H$. Since the angular momentum
quantum numbers, i.e. $l_{12},l_{3}$, do not appear explicitly in
our formalism, therefore the number of coupled equations which are
fixed according to the spin-isospin states are strongly reduced.
This is an indication that the present formalism automatically
considers all partial waves without any truncation on the space
part. Considering the spin-isospin degrees of freedom for both
$^{3}H$ and $^{3}He$ states yields the same number of coupled
equations and it leads to 8, 12, 16 and 24 coupled equations for
different combinations of the total spin-isospin states $S-T$:
$(\frac{1}{2}-\frac{1}{2})$, $(\frac{1}{2}-\frac{3}{2}^{\ast})$,
$(\frac{3}{2}^{\ast}-\frac{1}{2})$ and
$(\frac{3}{2}^{\ast}-\frac{3}{2}^{\ast})$ respectively. The star
superscript indicates all the spin or isospin states that we have
taken into account up to a specific value. It is clear that in the
3D formalism, e.g. for a fully charge dependent calculation, there
is only 24 coupled equations, whereas in the PW approach after
truncation of the Hilbert space to $T=\frac{1}{2}$ there is 42
coupled equations. Therefore our 3D formalism leads to a small
number of coupled equations in comparison with the very large number
of coupled equations in the truncated PW formalism. However, it
should be mentioned that our formulation leads to coupled equations
in three variables for the amplitudes, whereas the PW formulation
after truncation leads to a finite number of coupled equations in
two variables for the amplitudes. So the 3D formulation leads to a
lesser number of coupled integral equations in three dimensions and
the PW formulations leads to more coupled integral equations in two
dimensions. Thus, the price for the smaller number of equations is
the higher dimensionality of the integral equations. In other words,
algebraic simplification is achieved by a more involved numerical
scheme.

\section{Numerical Results for $^{3}$H} \label{sec:numerical results}

\subsection{Triton Binding Energy} \label{subsec:3H BE}

In order to be able to test our realistic 3D formalism for the 3N
bound state we solve the three-dimensional Faddeev integral
equations, Eq.~(\ref{eq.FC-magnitude}). We calculate the Triton
binding energy by solving eight coupled Faddeev equations for
$(\frac{1}{2}-\frac{1}{2})$ spin-isospin states and compare our
results with the other PW results. In this respect, we use Bonn
one-boson-exchange (OBE) potential in the parametrization of Bonn-B
\cite{Machleidt-ANP19} and in an operator form which can be
incorporated in the 3D formalism \cite{Fachruddin-PRC62}. In the
numerical treatment, the dependence of Faddeev components to the
continuous momentum and the angle variables, should be replaced by a
dependence on certain discrete values. For this purpose we use the
Gaussian quadrature grid points.

\begin{table}[hbt]
\caption {The calculated binding energy $E_{t}$ of the
three-dimensional Faddeev integral equations as function of the
number of the grid points in the Jacobi momenta $N_{jac}$ and the
spherical angles $N_{sph}$. The number of the grid points in polar
angles is twenty. The calculations are based on the Bonn-B
potential. }
\begin{tabular}{cccccccccccc}
\hline \hline
$N_{jac}$  &&&&&  $N_{sph}$ &&&&& $E_{t}$ [MeV]  \\
\hline
32  &&&&&  20 &&&&& -8.154  \\
32  &&&&&  24 &&&&& -8.153  \\
36  &&&&&  20 &&&&& -8.153  \\
36  &&&&&  24 &&&&& -8.152  \\
40  &&&&&  20 &&&&& -8.152  \\
40  &&&&&  24 &&&&& -8.152  \\
\hline \hline
\end{tabular}
\label{table:BE in 3D}
\end{table}

The coupled Faddeev equations represent a set of three dimensional
homogenous integral equations, which after discreatization turns
into a huge matrix eigenvalue equation. The huge matrix eigenvalue
equation requires an iterative solution method. We use a
Lanczos-like scheme that is proved to be very efficient for nuclear
few-body problems \cite{Stadler-PRC44}. The momentum variables have
to cover the interval $[0,\infty]$. In practice we limit the
intervals to suitable cut-offs and their values are chosen large
enough to achieve cut-off independence. The functional behavior of
the kernel of eigenvalue equation is determined by the
anti-symmetrized two-body $t-$matrix. We also solve the
Lippman-Schwinger equation for the fully-off-shell two-body
$t-$matrix in an approach based on a Helicity representation
directly as a function of the Jacobi vector variables (see appendix
\ref{app:t matrix}). For anti-symmetrized two-body $t-$matrix
calculations forty grid points for the Jacobi momentum variables,
thirty two grid points for the spherical angle variables and twenty
grid points for the polar angle variables have been used
respectively. Since the coupled integral equations require a very
large number of interpolations, we use the cubic Hermitian splines
of Ref. \cite{Huber-FBS22} for its accuracy and high computational
speed.

\begin{table}[hbt]
\caption {A list of Triton binding energy calculations ordered
according to $j_{12}^{max}$ by different authors using slightly
different numerical methods. All results for binding energies are
related to the total isospin $T=\frac{1}{2}$.}
\begin{tabular}{cccccccccccc}
\hline \hline
$j_{12}^{max}$  &&&&& Ref. &&&&& $E_{t}$ [MeV]  \\
\hline   1 &&&&& \\
 &&&&& \cite{Brandenburg-PRC37}  &&&&& -8.14 \\
 &&&&& \cite{Julia-Diaz-PRC65}, \cite{Valcarce-RPP68}  &&&&& -8.17 \\
 &&&&&  \cite{Schadow-FBS28}                &&&&& -8.165 \\
 &&&&& \cite{Sammarruca-PRC46}, \cite{Haidenbauer-PRC53}  &&&&& -8.16 \\
 \hline
2 &&&&& \\
&&&&& \cite{Schadow-NPA631}, \cite{Schadow-PRC63}  &&&&& -8.088 \\
&&&&& \cite{Adam-PRC69}  &&&&& -8.100 \\
&&&&& \cite{Schadow-NPA631}  &&&&& -8.101 \\
&&&&& \cite{Schadow-FBS28} &&&&& -8.103 \\
 \hline
 3 &&&&& \\
 &&&&& \cite{Glockle-PRL71}      &&&&& -8.14 \\
\hline
 4 &&&&& \\
&&&&& \cite{Machleidt-ANP19}, \cite{Li-PRC45}    &&&&& -8.13 \\
&&&&& \cite{Sammarruca-PRC46}, \cite{Witala-PRC43}  &&&&& -8.14 \\
\hline\hline
\end{tabular}
\label{table:BE in PW}
\end{table}

In Table \ref{table:BE in 3D} we show the convergence of the Triton
binding energy as function of the number of the grid points for
Bonn-B potential in the 3D approach. As demonstrated in this Table,
the calculation of Triton binding energy converges to a value of
$E_{t}=-8.152$ MeV. The results of the Faddeev equations with
different PW based methods are presented in Table \ref{table:BE in
PW} in order to compare them with our calculations. The overall
agreement is quite satisfactory. As we can see from this comparison
our result provides the same accuracy while the numerical procedure
is actually easier to implement.

\subsection{Expectation Value of
the Hamiltonian Operator} \label{subsec:test of calculations}

In this section we investigate the numerical stability of the
presented algorithm and the 3D formalism of the Faddeev equations.
With the binding energy $E_{t}$ and the Faddeev component
$|\psi\rangle $ available, we are able to calculate the total wave
function $|\Psi\rangle $ from Eq. (\ref{eq.WF}) by considering the
choice of coordinate system which is used in representation of Eq.
(\ref{eq.FC-magnitude}). So we can evaluate the expectation value of
the Hamiltonian operator $H$ and compare this value to the
previously calculated binding energy of the eigenvalue equation.
Explicitly we evaluate the following expression:
\begin{eqnarray}
\langle \Psi |H| \Psi \rangle =  \langle \Psi |H_0| \Psi \rangle
      +    \langle \Psi | V | \Psi \rangle = 3 \,\langle \psi
 |H_0| \Psi \rangle + 3 \,\langle \Psi | V_{12} | \Psi \rangle,
 \label{eq.Hamiltonian}
\end{eqnarray}
where
\begin{eqnarray}
\langle \psi |H_0| \Psi \rangle &=& \sum_{\alpha} \int d^{3}p \int
d^{3}q \, \sum_{\alpha'}  \int d^{3}p'  \int d^{3}q' \,\, \langle \,
\psi | \, {\bf p}\,{\bf q} \, \alpha  \rangle \, \langle \, {\bf
p}\,{\bf q} \, \alpha |H_0| \, {\bf p'}\,{\bf q'} \, \alpha' \rangle
\, \langle \, {\bf p'}\,{\bf q'} \, \alpha' | \Psi \rangle \nonumber
\\*
 &=&  \int_{0}^{\infty} dp \, p^{2} \int_{0}^{\infty} dq \, q^{2}
\,\, \left ( \frac{p^{2}}{m}+\frac{3q^{2}}{4m} \right )
 \int_{-1}^{+1} dx_{p} \int_{0}^{2\pi} d\varphi_{p} \int_{-1}^{+1}
dx_{q} \int_{0}^{2\pi} d\varphi_{q}  \nonumber \\*
 && \times \, \sum_{\alpha}  \psi^{\alpha}(p \,\, q \,\, x_{pq})
\,\, \Psi^{\alpha}(p \,\, q \,\, x_{pq}),
  \label{eq.H0}
\end{eqnarray}

\begin{eqnarray}
\langle \Psi |V_{12}| \Psi \rangle &=& \sum_{\alpha} \int d^{3}p
\int d^{3}q \, \sum_{\alpha'}  \int d^{3}p'  \int d^{3}q' \,\,
\langle \, \Psi | \, {\bf p}\,{\bf q} \, \alpha  \rangle \, \langle
\, {\bf p}\,{\bf q} \, \alpha |V_{12}| \, {\bf p'}\,{\bf q'} \,
\alpha' \rangle \, \langle \, {\bf p'}\,{\bf q'} \, \alpha' | \Psi
\rangle. \nonumber \\* \label{eq.V12}
\end{eqnarray}

As is well known, the rotational, parity and time-reversal
invariance restricts any NN potential $V_{12}$ to be formed out of
six independent terms \cite{Wolfenstein-PR96}, as
\begin{eqnarray}
V_{12} ({\bf p} \,, {\bf p'}) = \langle \, {\bf p} |V_{12}| \, {\bf
p'} \rangle = \sum _{i=1} ^{6}  v_{i} (p \,, p' \,, x_{pp'}) \,\,
W_{i}, \label{eq.V12 operator}
\end{eqnarray}
here $v_{i} (p \,, p' \,, x_{pp'})$ are scalar spin-independent
functions, which depend on the magnitudes of the Jacobi momenta
${\bf p} \,, {\bf p'}$ and the angle between them, $x_{pp'}\equiv
\hat{{\bf p}}.\hat{{\bf p}}'$, and $W_{i}$ (i=1 to 6) are operators
to the spin states of the two-nucleon such that
\begin{eqnarray}
V_{12} \, _{m_{s_{1}}  m_{s_{2}}  m'_{s_{1}}  m'_{s_{2}}} ({\bf p}
\,, {\bf p'}) &=& \langle \, {\bf p} \, m_{s_{1}}  m_{s_{2}}
|V_{12}| \, {\bf p'} \, m'_{s_{1}}  m'_{s_{2}} \rangle  \nonumber
\\* &=& \sum
_{i=1} ^{6} v_{i} (p \,, p' \,, x_{pp'}) \,\, \langle \, m_{s_{1}}
m_{s_{2}} |W_{i}| \,  m'_{s_{1}} m'_{s_{2}} \rangle,  \label{eq.V12
operator spin}
\end{eqnarray}
so the matrix elements of NN potential can be evaluated as:
\begin{eqnarray}
\langle \, {\bf p}\,{\bf q} \, \alpha |V_{12}| \, {\bf p'}\,{\bf q'}
\, \alpha' \rangle &=& \delta^{3}({\bf q}-{\bf q'})  \, \langle  \,
\alpha_{T} |V_{12}^{T}| \, \alpha'_{T} \rangle \, \langle \, {\bf p}
\, \alpha_{S} |V_{12}| \, {\bf p'} \, \alpha'_{S} \rangle,
\label{eq.V12 spin isospin}
\end{eqnarray}
where $V_{12}^{T}$ is the isospin part of the potential, it is unity
for the isospin-independent terms and $\tau_{1}.\tau_{2}$ for the
isospin-dependent terms. So it can be easily evaluated as
\begin{eqnarray}
\langle  \, \alpha_{T} |V_{12}^{T}| \, \alpha'_{T} \rangle =
\textbf{T} \, \delta_{\alpha_{T} \alpha'_{T}}  , \,\, \textbf{T}=\left\{%
\begin{array}{ll}
    1, & \hbox{isospin-independent terms;} \\
    2t_{12}^{2}-3, & \hbox{isospin-dependent terms.} \\
\end{array}%
\right.     \label{eq.V12 isospin}
\end{eqnarray}

The spin-space part of the potential can be evaluated as:
\begin{eqnarray}
  \langle \, {\bf p} \, \alpha_{S} |V_{12}| \, {\bf p'} \,
\alpha'_{S} \rangle &=& \sum_{\gamma_{S}} \sum_{\gamma'_{S}} \langle
\, \alpha_{S} |  \, \gamma_{S} \rangle \, \langle \, \gamma'_{S}
 | \, \alpha'_{S} \, \rangle  \langle \, {\bf p} \, \gamma_{S}
|V_{12}| \, {\bf p'} \, \gamma'_{S}
 \rangle \nonumber
\\* &=& \sum_{\gamma_{S}} \sum_{\gamma'_{S}}
g_{\alpha \gamma}^{S} \, g_{\alpha' \gamma'}^{S}  \langle \, {\bf p}
\, \gamma_{S} |V_{12}| \, {\bf p'} \, \gamma'_{S}
 \rangle   \nonumber
\\* &=& \sum_{\gamma_{S}} \sum_{\gamma'_{S}}
g_{\alpha \gamma}^{S} \, g_{\alpha' \gamma'}^{S}  \, \delta_{
m_{s_{3}} m'_{s_{3}}} \langle \, {\bf p} \, m_{s_{1}}
m_{s_{2}}|V_{12}| \, {\bf p'} \, m'_{s_{1}} m'_{s_{2}}
 \rangle  \nonumber
\\* &=& \sum_{\gamma_{S}} \sum_{\gamma'_{S}}
g_{\alpha \gamma}^{S} \, g_{\alpha' \gamma'}^{S}  \, \delta_{
m_{s_{3}} m'_{s_{3}}} \, V_{12} \, _{m_{s_{1}}  m_{s_{2}} m'_{s_{1}}
m'_{s_{2}}} ({\bf p} \,, {\bf p'}), \label{eq.V12 spin}
\end{eqnarray}
inserting Eqs. (\ref{eq.V12 isospin}) and (\ref{eq.V12 spin}) into
Eq. (\ref{eq.V12 spin isospin}) yields:
\begin{eqnarray}
\langle \, {\bf p}\,{\bf q} \, \alpha |V_{12}| \, {\bf p'}\,{\bf q'}
\, \alpha' \rangle &=& \delta^{3}({\bf q}-{\bf q'})  \, \textbf{T}
\, \delta_{\alpha_{T} \alpha'_{T}}  \, \sum_{\gamma_{S}}
\sum_{\gamma'_{S}} g_{\alpha \gamma}^{S} \, g_{\alpha' \gamma'}^{S}
\, \delta_{ m_{s_{3}} m'_{s_{3}}} \, V_{12} \, _{m_{s_{1}} m_{s_{2}}
m'_{s_{1}}  m'_{s_{2}}} ({\bf p} \,, {\bf p'}), \nonumber
\\* \label{eq.V12 final}
\end{eqnarray}
by these considerations the expectation value of the NN potential,
Eq. (\ref{eq.V12}), can be rewritten as:
\begin{eqnarray}
\langle \Psi |V_{12}| \Psi \rangle &=& \sum_{\alpha} \sum_{\alpha'}
\textbf{T} \, \delta_{\alpha_{T} \alpha'_{T}} \, \sum_{\gamma_{S}}
\sum_{\gamma'_{S}} g_{\alpha \gamma}^{S} \, g_{\alpha' \gamma'}^{S}
\, \delta_{ m_{s_{3}} m'_{s_{3}}} \, \nonumber
\\* && \hspace{-17mm} \times
\int_{0}^{\infty} dp \, p^{2}  \int_{-1}^{+1} dx_{p} \int_{0}^{2\pi}
d\varphi_{p} \int_{0}^{\infty} dp' \, p'^{2} \int_{-1}^{+1} dx'_{p}
\int_{0}^{2\pi} d\varphi'_{p}  V_{12} \, _{m_{s_{1}} m_{s_{2}}
m'_{s_{1}} m'_{s_{2}}} (p \,, p' \,, x_{pp'})    \nonumber
\\* &&  \hspace{-17mm} \times \int_{0}^{\infty} dq \, q^{2}  \int_{-1}^{+1} dx_{q}
\int_{0}^{2\pi} d\varphi_{q} \,\,  \Psi^{\alpha}(p \,\, q \,\,
x_{pq}) \,\, \Psi^{\alpha'}(p' \,\, q \,\, x_{p'q}),  \label{eq.V12
expectation}
\end{eqnarray}
where $x_{pp'}\equiv \hat{{\bf p}}.\hat{{\bf p}}' =x_{p}x_{p'}+
\sqrt{1-x_{p}^{2}}\sqrt{1-x_{p'}^{2}}\sin (\phi_{p}-\phi_{p'})$ and
$x_{p'q}\equiv \hat{{\bf p}}'.\hat{{\bf q}} =x_{p'}x_{q}+
\sqrt{1-x_{p'}^{2}}\sqrt{1-x_{q}^{2}}\sin (\phi_{p'}-\phi_{q})$.

\begin{table}[hbt]
\caption{The expectation values of the kinetic energy $\langle
H_{0}\rangle$, the NN interaction $\langle V\rangle$ and the
Hamiltonian operator $\langle H\rangle$ calculated in the 3D scheme
as a function of the number of the grid points in the Jacobi momenta
$N_{jac}$ and the spherical angles $N_{sph}$ for the Triton. The
number of the grid points in polar angles is twenty. The
calculations are based on the Bonn-B potential. Additionally the
expectation values of the Hamiltonian operator are compared with the
Triton binding energy results from the three-dimensional Faddeev
integral equations. All energies are given in MeV. }
\begin{tabular} {cccccccccccccccc}
\hline\hline
$N_{jac}$ &&& $N_{sph}$ &&& $\langle H_0\rangle$ &&& $\langle V\rangle$ &&& $\langle H\rangle$ &&& $E_{t}$\\
\hline
32  &&& 20 &&& +39.222 &&& -47.356 &&& -8.134 &&& -8.154  \\
32  &&& 24 &&& +39.222 &&& -47.356 &&& -8.134 &&& -8.154  \\
36  &&& 20 &&& +39.222 &&& -47.357 &&& -8.135 &&& -8.153  \\
36  &&& 24 &&& +39.222 &&& -47.357 &&& -8.135 &&& -8.152  \\
40  &&& 20 &&& +39.223 &&& -47.358 &&& -8.135 &&& -8.152  \\
40  &&& 24 &&& +39.223 &&& -47.358 &&& -8.135 &&& -8.152  \\
 \hline\hline
\end{tabular} \label{table:expectation values}
\end{table}

The expectation values of the kinetic energy $\langle H_0\rangle$,
the two-body interaction $\langle V\rangle$ and the Hamiltonian
operator $\langle H \rangle$ are listed in Table
\ref{table:expectation values} for Bonn-B potential calculated in
the 3D scheme as a function of the number of the grid points in the
Jacobi momenta $N_{jac}$ and the spherical angles $N_{sph}$. In the
same Table, the Triton binding energies calculated in the 3D scheme
are also shown in order to compare with the expectation values of
the Hamiltonian operator. One can see that the energy expectation
value and the eigenvalue energies $E_{t}$ agree with good accuracy.

\section{Summary and Outlook}\label{sec:summary}
In this paper we have introduced the three-dimensional Faddeev
integral equations for the calculation of the Triton binding energy
with the spin-isospin dependent potential. In comparison with the PW
approach, as is commonly used, this direct approach has greater
advantages. The pertinent results can be summarized as follows:

   1) The 3D formalism leads only to a strictly finite number of
coupled three-dimensional integral equations to be solved, whereas
in the PW case after truncation one has a set of finite number of
coupled equations with kernels containing relatively complicated
geometrical expressions. So the 3D formalism avoids the highly
involved angular momentum algebra occurring for the permutations and
also automatically consider all the partial waves without any
truncation on the space part. However the 3D formulation leads to a
lesser number of coupled integral equations in three dimensions and
the PW formulations leads to more coupled integral equations in two
dimensions.

    2) Our result for the Triton binding energy with Bonn-B potential
is in good agreement with the pervious values calculated with the
standard PW approach. The stability of present algorithm and the 3D
formalism of Faddeev components as presented in this paper have been
achieved with the calculation of the expectation value of the
Hamiltonian operator and we have reached to a resonable agreement
between the obtained energy eigenvalue and expectation value of the
Hamiltonian operator. The 3N bound state calculations with AV18
potential is also potentially valuable and the numerical results
with this potential will be reported in the future.

    3) We predict that the incorporation of three-nucleon force probably
    will be less cumbersome in a realistic 3D approach. This
is very promising and nourishes our hope that four-nucleon bound
state formulation and calculations with realistic two and
three-nucleon forces in a realistic 3D approach will be more easily
implemented than the traditional partial wave based method.

The calculations of three-nucleon bound state, with the
phenomenological Tucson-Melbourne (TM) $2\pi$ exchange three nucleon
potential, and the formulation of the four-nucleon bound state is
currently underway and they will be reported before long
\cite{Hadizadeh-preparation}.

\section*{Acknowledgments}

We would like to thank W. Gl\"{o}ckle, Ch. Elster and I. Fachruddin
for using their helicity formalism related to the two-body
$t$-matrix. This work was supported by the research council of the
University of Tehran.

\appendix

\section{$g_{\gamma \alpha }$ Clebsch-Gordan coefficients} \label{app:CG coeeficients}
In the usual coupling scheme, for the three identical particles with
spin $\frac{1}{2}$, in order to completely classify the states of
definite total spin the quantum numbers
\begin{equation}
| \, \gamma_{S} \, \rangle \equiv | \, s_{1} m_{s_{1}} \, s_{2}
m_{s_{2}} \, s_{3} m_{s_{3}} \, \rangle \equiv |  m_{s_{1}} \,
m_{s_{2}} \, m_{s_{3}} \rangle, \label{eq.free basis}
\end{equation}
are replaced by the set
\begin{eqnarray}
  | \, \alpha_{S} \, \rangle &\equiv& | \, ((s_{1}\,\,s_{2})s_{12}
 \,\, s_{3}) S \, M_{S} \, \rangle \equiv | \, (s_{12} \,\,
 \frac{1}{2}) S \, M_{S} \, \rangle.  \label{eq.real basis}
\end{eqnarray}

The 3N basis states $ | \, \alpha_{S} \, \rangle$ can be obtained
from free 3N basis states $| \gamma_{S} \rangle$ as:
\begin{eqnarray}
\left\{%
\begin{array}{ll}
    | \, (1 \,\, \frac{1}{2}) \frac{3}{2} \, + \frac{3}{2} \,
    \rangle \equiv  |  \uparrow \,
\uparrow \, \uparrow \rangle \\
   | \, (1 \,\, \frac{1}{2})
\frac{3}{2} \, + \frac{1}{2} \,
   \rangle \equiv \frac{1}{\sqrt{3}} \biggl \{ | \downarrow \,
   \uparrow \, \uparrow \rangle + | \uparrow \, \downarrow \,
   \uparrow \rangle + | \uparrow \,
\uparrow \, \downarrow \rangle \biggr \} \\
| \, (1 \,\, \frac{1}{2}) \frac{3}{2} \, - \frac{1}{2} \, \rangle
\equiv \frac{1}{\sqrt{3}} \biggl \{ | \uparrow \, \downarrow \,
\downarrow \rangle + | \downarrow \, \uparrow \, \downarrow \rangle
+ | \downarrow \,
\downarrow \, \uparrow \rangle \biggr \} \\
 | \, (1 \,\, \frac{1}{2}) \frac{3}{2} \, - \frac{3}{2} \, \rangle
 \equiv  |  \downarrow \,
\downarrow \, \downarrow \rangle \\
| \, (1 \,\, \frac{1}{2}) \frac{1}{2} \, + \frac{1}{2} \, \rangle
\equiv \frac{1}{\sqrt{6}} \biggl \{ | \uparrow \, \downarrow \,
\uparrow \rangle + | \uparrow \, \uparrow \, \downarrow \rangle -2 |
\downarrow \,
\uparrow \, \uparrow \rangle \biggr \} \\
| \, (1 \,\, \frac{1}{2}) \frac{1}{2} \, - \frac{1}{2} \, \rangle
\equiv \frac{1}{\sqrt{6}} \biggl \{ | \downarrow \, \uparrow \,
\downarrow \rangle + | \downarrow \, \downarrow \, \uparrow \rangle
-2 | \uparrow \,
\downarrow \, \downarrow \rangle \biggr \} \\
| \, (0 \,\, \frac{1}{2}) \frac{1}{2} \, + \frac{1}{2} \, \rangle
\equiv \frac{1}{\sqrt{2}} \biggl \{ | \uparrow \, \uparrow \,
\downarrow \rangle - | \uparrow \, \downarrow \,
\uparrow \rangle \biggr \} \\
| \, (0 \,\, \frac{1}{2}) \frac{1}{2} \, - \frac{1}{2} \, \rangle
\equiv \frac{1}{\sqrt{2}} \biggl \{ | \downarrow \, \uparrow \,
\downarrow \rangle - | \downarrow \, \downarrow \,
\uparrow \rangle \biggr \} \\
\end{array}%
\right. \label{eq.real basis}
\end{eqnarray}

 If one considers all total spin states, i.e. $S=\frac{1}{2}$ and
$S=\frac{3}{2}$, the relevant Clebsch-Gordan coefficients $g_{\gamma
\alpha }^{S}$ are $1, \frac{1}{\sqrt{3}}, \frac{1}{\sqrt{6}},
-\sqrt{\frac{2}{3}}, \frac{1}{\sqrt{2}}, -\frac{1}{\sqrt{2}}$. As
indicated in section \ref{subsec:discussion} the isospin states are
similar to the spin states, but the third component of total
isospins is restricted to $M_{T}=\frac{+1}{2}$ for $^{3}He$ and
$M_{T}=\frac{-1}{2}$ for $^{3}H$. Thus for a fully charge dependent
calculation the necessary isospin coefficients $g_{\gamma \alpha
}^{T}$ are $\frac{1}{\sqrt{3}}, \frac{1}{\sqrt{6}},
-\sqrt{\frac{2}{3}}, \frac{1}{\sqrt{2}}, -\frac{1}{\sqrt{2}}$. Since
in our calculations for the Triton binding energy we consider only
the total spin-isospin states $(S-T)=(\frac{1}{2}-\frac{1}{2})$,
therefore we only use the following Clebsch-Gordan coefficients $
\frac{1}{\sqrt{6}}, -\sqrt{\frac{2}{3}}, \frac{1}{\sqrt{2}},
-\frac{1}{\sqrt{2}}$.

\section{Anti-Symmetrized NN $T$-matrix and Connection to Helicity
Representation} \label{app:t matrix}

 In our formulation of the 3N bound state, we need the physical
representation of NN $t$-matrix or matrix elements $_{a}\langle{\bf
p} \, m_{s_{1}} m_{s_{2}} \, m_{t_{1}} m_{t_{2}} |t(\varepsilon)|
{\bf p}' \, m'_{s_{1}} m'_{s_{2}} \, m'_{t_{1}} m'_{t_{2}} \rangle
_{a}$. The connection of these matrix elements to those in the
momentum-helicity basis is given in Ref.~\cite{Fachruddin-PRC68},
here we prepare this connection according to the notation to be used
in our work. First, we introduce the momentum-helicity basis states
for the total spin $s_{12}$ and the relative momentum ${\bf p}$ of
the two nucleons as:
\begin{equation}
 |{\bf p};\hat{{\bf p}}s_{12}\lambda \rangle,
\end{equation}
where $\lambda$ is the eigenvalue of the helicity operator ${\bf
s_{12}.\hat{{\bf p}}} $. By introducing parity operator $P$ and the
two-nucleon isospin states $|t_{12} \, m_{t_{12}}\rangle$, the
anti-symmetrized two-nucleon basis states are given as:
\begin{equation}
|{\bf p};\hat{{\bf p}}s_{12}\lambda ;t_{12} \rangle ^{\pi a}
 \equiv \frac {1}{\sqrt{2}}
     (1-\eta_{\pi}(-)^{s_{12}+t_{12}}) \, |t_{12}\rangle \, |{\bf
p};\hat{{\bf p}}s_{12}\lambda \rangle _{\pi},
\end{equation}
with the parity eigenvalues $\eta_{\pi}=\pm1$ and eigenstates $|{\bf
p};\hat{{\bf p}}s_{12}\lambda \rangle _{\pi}=\frac{1}{\sqrt{2}}
(1+\eta_{\pi} P) |{\bf p};\hat{{\bf p}}s_{12}\lambda \rangle$. Based
on these basis states the NN $t$-matrix element is defined as:
\begin{equation}
t_{\lambda \lambda '}^{\pi s_{12}t_{12}}({{\bf p}, \, {\bf p}'
};\varepsilon) \equiv \, ^{\pi a} \langle {\bf p};\hat{{\bf
p}}s_{12}\lambda ;t_{12} | t(\varepsilon)|{\bf p}';\hat{{\bf
p}}'s_{12}\lambda' ;t_{12} \rangle ^{\pi a}.
\end{equation}

As shown in Ref.~\cite{Fachruddin-PRC68}, the selection of ${\bf
p}'$ parallel to the $z$-axis allows, together with the properties
of the potential, that the angular dependencies of the NN $t$-matrix
elements can be simplified as:
\begin{eqnarray}
t_{\lambda \lambda '}^{\pi s_{12}t_{12}}({{\bf p}, \, {\bf p}'
};\varepsilon) &=& e^{-i \lambda \Omega_{pp'} } \,\, t_{\lambda
\lambda '}^{\pi s_{12}t_{12}}({p{\bf \hat{n}}_{pp'}, \, p'{\bf
\hat{z}} };\varepsilon)  \nonumber
\\* &=&
e^{-i \lambda \Omega_{pp'} } \,\, e^{i \lambda' \phi_{pp'} } \,
t_{\lambda \lambda '}^{\pi s_{12}t_{12}}(p, p', \cos \theta_{pp'}
;\varepsilon) \nonumber
\\* &\equiv&
 e^{i(\lambda '\phi_{pp'} -\lambda \Omega_{pp'} )} \,\, t_{\lambda
\lambda '}^{\pi s_{12}t_{12}}(p,p',\cos \theta_{pp'} ;\varepsilon),
\end{eqnarray}
the direction ${\bf \hat{n}}_{pp'}$ can be determined by the
spherical and polar angles $\vartheta _{pp'}$ and $\varphi_{pp'}$,
where
\begin{eqnarray}
\cos \theta_{pp'} &=&  \cos \theta_{p} \cos \theta_{p'} +\sin
\theta_{p} \sin \theta_{p'} \cos (\phi_{p} -\phi_{p'}), \nonumber
\\*
\sin \theta_{pp'} e^{i \varphi_{pp'}} &=&  -\cos \theta_{p} \sin
\theta_{p'} +\sin \theta_{p} \cos \theta_{p'} \cos (\phi_{p}
-\phi_{p'}) + i \sin \theta_{p} \sin (\phi_{p} -\phi_{p'}),
\label{eq.cos-teta}
\end{eqnarray}
and the exponential factor $ e^{i(\lambda' \phi_{pp'} -\lambda
\Omega )} $ is calculated as:
\begin{eqnarray}
e^{i\lambda \Omega_{pp'} }  &=& \frac{\sum ^{s_{12}}_{N=-s_{12}}
D^{s_{12}}_{N\lambda }( \phi_{p} \, \theta_{p} \, 0 )
D^{s_{12}}_{N\lambda '} (\phi_{p'} \, \theta_{p'} \, 0
)}{D^{s_{12}}_{\lambda '\lambda }( \phi_{pp'} \, \theta_{pp'} \,
0)}, \nonumber
\\*
e^{i(\lambda '\phi_{pp'} -\lambda \Omega_{pp'} )}  &=& \frac{\sum
^{s_{12}}_{N=-s_{12}}e^{iN(\phi_{p} -\phi_{p'}
)}d^{s_{12}}_{N\lambda }(\theta_{p} )d^{s_{12}}_{N\lambda '}
(\theta_{p'} )}{d^{s_{12}}_{\lambda '\lambda }(\theta_{pp'})}.
\label{eq.exp-factor}
\end{eqnarray}

In the above expressions, $D^{s_{12}}_{N\lambda }( \phi_{p} \,
\theta_{p} \, 0 )$ are the Wigner D-functions and
$d^{s_{12}}_{\lambda '\lambda }(\theta )$ are rotation matrices
\cite{rose}. Finally the connection of the $t$-matrix elements $
_{a}\langle{\bf p} \, m_{s_{1}} m_{s_{2}} \, m_{t_{1}} m_{t_{2}}
|t(\varepsilon)| {\bf p}' \, m'_{s_{1}} m'_{s_{2}} \, m'_{t_{1}}
m'_{t_{2}} \rangle _{a}$ to those in the momentum-helicity basis,
namely $t_{\lambda \lambda '}^{\pi s_{12}t_{12}}({\bf p},{\bf
p}';\varepsilon)$, is given as:
\begin{eqnarray}
&& _{a}\langle{\bf p} \, m_{s_{1}} m_{s_{2}} \, m_{t_{1}} m_{t_{2}}
|t(\varepsilon)| {\bf p}' \, m'_{s_{1}} m'_{s_{2}} \, m'_{t_{1}}
m'_{t_{2}} \rangle _{a}  \nonumber \\* && \hspace{40mm} =\frac{1}{4}
\, \delta _{m_{t_{1}} +m_{t_{2}},m'_{t_{1}}+m'_{t_{2}}}
e^{-i(\lambda _{0}\phi _p -\lambda _{0}'\phi _{p' })} \, \sum
_{s_{12}\pi t_{12}}( 1-\eta _{\pi }(-)^{s_{12}+t_{12}}) \nonumber
\\* && \hspace{40mm} \times
C(\frac{1}{2}\frac{1}{2}t_{12}; m_{t_{1}} m_{t_{2}} )
C(\frac{1}{2}\frac{1}{2}t'_{12}; m'_{t_{1}} m'_{t_{2}} ) \nonumber
\\* && \hspace{40mm} \times C(\frac{1}{2}\frac{1}{2}s_{12}; m_{s_{1}} m_{s_{2}} )
C(\frac{1}{2}\frac{1}{2}s'_{12}; m'_{s_{1}} m'_{s_{2}} ) \nonumber
\\* && \hspace{40mm} \times
\sum _{\lambda \lambda '}d^{s_{12}}_{\lambda _{0}\lambda }(\theta
_p)d^{s_{12}}_{\lambda _{0}'\lambda '}(\theta _{p' })t_{\lambda
\lambda '}^{\pi s_{12}t_{12}}({{\bf p}, \, {\bf p}' };\varepsilon).
\label{eq.t_a-t_helicity}
\end{eqnarray}

It should be mentioned that $t_{\lambda \lambda '}^{\pi
s_{12}t_{12}}(p,p',\cos \theta_{pp'} ;\varepsilon)$ obeys a set of
coupled Lippman-Schwinger equations which for \(S = 0\) it is a
single equation but for \(S = 1\) it is a set of two coupled
equations (Ref.~\cite{Fachruddin-PRC62}). So the matrix elements of
the anti-symmetrized NN $t$-matrix, which explicitly appears in Eq.
(\ref{eq.kernel}), is functionally the same as Eq.
(\ref{eq.t_a-t_helicity}) and can be obtained as:
\begin{eqnarray}
 t_{a}\, _{\, m_{s_{1}} m_{s_{2}} \, m_{t_{1}} m_{t_{2}}} ^{\,
 m'_{s_{2}} m'_{s_{3}} \, m'_{t_{2}} m'_{t_{3}}}(p,\tilde{\pi},
 x_{p \tilde{\pi}} ;\epsilon) &\equiv& \, _{a}\langle{\bf p} \,
 m_{s_{1}} m_{s_{2}} \, m_{t_{1}} m_{t_{2}} |t(\varepsilon)| {\bf
 \pi} \, m'_{s_{1}} m'_{s_{2}} \, m'_{t_{1}} m'_{t_{2}} \rangle
 _{a} \nonumber \\* &=&  \frac{1}{4} \, \delta _{m_{t_{1}}
 +m_{t_{2}},m'_{t_{1}}+m'_{t_{2}}} e^{-i(\lambda _{0}\phi _p
 -\lambda _{0}'\phi _{\tilde{\pi} })} \, \sum _{s_{12}\pi t_{12}}(
 1-\eta _{\pi }(-)^{s_{12}+t_{12}}) \nonumber
\\* &\times&
C(\frac{1}{2}\frac{1}{2}t_{12}; m_{t_{1}} m_{t_{2}} )
C(\frac{1}{2}\frac{1}{2}t'_{12}; m'_{t_{1}} m'_{t_{2}} ) \nonumber
\\* &\times&   C(\frac{1}{2}\frac{1}{2}s_{12}; m_{s_{1}} m_{s_{2}} )
C(\frac{1}{2}\frac{1}{2}s'_{12}; m'_{s_{1}} m'_{s_{2}} ) \nonumber
\\* &\times&
\sum _{\lambda \lambda '}d^{s_{12}}_{\lambda _{0}\lambda }(\theta
_p)d^{s_{12}}_{\lambda _{0}'\lambda '}(\theta _{\tilde{\pi}
})t_{\lambda \lambda '}^{\pi s_{12}t_{12}}({{\bf p}, \, {\bf
\tilde{\pi}} };\varepsilon), \label{eq.t_a-final}
\end{eqnarray}
with the same variables as Eqs. (\ref{eq.cos-teta}) and
(\ref{eq.exp-factor}) with $\tilde{\pi}, \theta _{\tilde{\pi}}, \phi
_{\tilde{\pi}}$ instead of $p', \theta _{p'}, \phi _{p'}$.


\begin{thebibliography}{0}

\bibitem{Hiyama-PRL85} E. Hiyama et al., {\it Phys. Rev. Lett.} {\bf 85}, 270 (2000).

\bibitem{Usukura-PRB59} J. Usukura, K. Varga and Y. Suzuki, {\it Phys. Rev.} {\bf B59}, 5652 (1999).

\bibitem{Viviani-PRC71} M. Viviani, A. Kievsky and S. Rosati, {\it Phys. Rev.} {\bf C71}, 024006 (2005).

\bibitem{Viringa-PRC62} R. B. Viringa et al., {\it Phys. Rev.} {\bf C62}, 014001 (2000).

\bibitem{Navratil-PRC62} P. Navr\'{a}til, J. P. Vary and B. R. Barret, {\it Phys. Rev. } {\bf C62}, 054311 (2000).

\bibitem{Barnea-PRC67} N. Barnea, W. Leidemann and G. Orlandini, {\it Phys. Rev.} {\bf C67}, 054003 (2003).

\bibitem{Sammarruca-PRC46} F. Sammarruca, D. P. Xu and R. Machleidt, {\it Phys. Rev.} {\bf C46}, 1636 (1992).

\bibitem{Song-AIP334} Y. Song and R. Machleidt, {\it AIP Conference Proceedings} {\bf 334}, 455
(1995).

\bibitem{Stadler-PRC51} A. Stadler, J. Adam Jr., H. Henning and P.U. Sauer {\it Phys. Rev.} {\bf C51}, 2896 (1995).

\bibitem{Nogga-PLB409} A. Nogga, D. H\"{u}ber, H. Kamada and W. Gl\"{o}ckle, {\it Phys. lett.} {\bf B409}, 19 (1997).

\bibitem{Stadler-PRL78} A. Stadler and F. Gross {\it Phys. Rev. Lett.} {\bf 78}, 26 (1997).

\bibitem{Fachruddin-PRC69} I. Fachruddin, W. Gl\"{o}ckle, Ch. Elster, A. Nogga, {\it Phys. Rev.} {\bf C69}, 064002 (2004).

\bibitem{Chen-PRL55} C. R. Chen, G.
L. Payne, J. L. Friar and B. F. Gibson, {\it Phys. Rev. Lett.}
{\bf 55}, 374 (1985)

\bibitem{Chen-PRC33} C. R. Chen, G. L. Payne,
J. L. Friar and B. F. Gibson, {\it Phys. Rev. } {\bf C33}, 1740
(1986)

\bibitem{Friar-PRC36} J. L. Friar, B. F. Gibson and G. L. Payne, {\it Phys. Rev. } {\bf C36},
1138 (1987)

\bibitem{Schellingerhout-PRA40} N. W.
Schellingerhout, L. P. Kok and G. D. Bosveld, {\it Phys. Rev. }
{\bf A40}, 5568 (1989)

\bibitem{Schellingerhout-PRC46} N. W.
Schellingerhout, J. J. Schut and L. P. Kok, {\it Phys. Rev. } {\bf
C46}, 1192 (1992)

\bibitem{Friar-PLB311} J. L. Friar, G. L. Payne, V. G. J. Stoks and J. J. de
Swart, {\it Phys. Lett.} {\bf B311}, 4 (1993).

\bibitem{Bedaque-NPA676} P. F. Bedaque, H.-W. Hammer and U. van Kolck {\it Nucl. Phys.} {\bf A676}, 357 (2000).

\bibitem{Epelbaum-PRL86} E. Epelbaum et al., {\it Phys. Rev. Lett.} {\bf 86},
4787 (2001).

\bibitem{Epelbaum-PRC66} E. Epelbaum et al., {\it Phys. Rev. } {\bf C66},
064001 (2002).

\bibitem{Platter-PLB607} L. Platter, H. -W. Hammer and U. -G. Mei{\ss}ner, {\it Phys. Lett.} {\bf B607}, 254 (2005).

\bibitem{Elster-FBS27} Ch. Elster, W. Schadow, A. Nogga, W. Gl\"{o}ckle, {\it Few-Body Systems} {\bf 27}, 83 (1999).

\bibitem{Liu-FBS33} H. Liu, Ch. Elster, W. Gl\"{o}ckle, {\it Few-Body Systems} {\bf 33},
241 (2003).

\bibitem{Hadizadeh-WS} M. R. Hadizadeh and S. Bayegan, (World
Scientific, Singapore, 2007, p. 16). {\it arXiv:nucl-th/0605067}.

\bibitem{Hadizadeh-FBS40} M. R. Hadizadeh and S. Bayegan, {\it Few-Body Systems} {\bf 40},
171 (2007). {\it arXiv:nucl-th/0605063}.

\bibitem{Hadizadeh-EPJA} M. R. Hadizadeh and S. Bayegan, {\it Eur. Phys. J.} {\bf A36},
201 (2008). {\it arXiv:0704.2056}.

\bibitem{Bayegan-EFB20} S. Bayegan, M. R. Hadizadeh and M. Harzchi, {\it to appear in Few Body Systems}. {\it arXiv:0711.4036}.

\bibitem{Fachruddin-PRC62} I. Fachruddin, Ch. Elster, and W. Gl\"{o}ckle, {\it Phys. Rev.} {\bf C62}, 044002 (2000).

\bibitem{Fachruddin-PRC68} I. Fachruddin, Ch. Elster, and W.
Gl\"{o}ckle, {\it Phys. Rev.} {\bf C68}, 054003 (2003).

\bibitem{rose} M. E. Rose, \textit{Elementary Theory of Angular Momentum} (Wiley, New York, 1957).

\bibitem{Machleidt-ANP19} R. Machleidt, {\it Adv. Nucl. Phys.} {\bf 19}, 189 (1989).

\bibitem{Stadler-PRC44} A. Stadler, W. Gl\"{o}ckle and P. U. Sauer, {\it Phys. Rev.} {\bf C44}, 2319 (1991).

\bibitem{Huber-FBS22} D. H\"{u}ber, H. Witala, A. Nogga, W. Gl\"{o}ckle and H. Kamada, {\it Few-Body Systems} {\bf 22}, 107 (1997).

\bibitem{Brandenburg-PRC37} R. A. Brandenburg, G. S. Chulick, R. Machleidt, A. Picklesimer, and R. M. Thaler, {\it Phys. Rev.} {\bf C37}, 1245 (1998).

\bibitem{Julia-Diaz-PRC65} B. Juli\'{a}-D\'{i}az, J. Haidenbauer, A. Valcarce, and F. Fern\'{a}ndez, {\it Phys. Rev.} {\bf C65}, 034001 (2002).

\bibitem{Valcarce-RPP68} A. Valcarce, H. Garcilazo, F. Fern\'{a}ndez, and P. Gonz\'{a}lez, {\it Rep. Prog. Phys.} {\bf 68}, 965 (2005).

\bibitem{Schadow-FBS28} W. Schadow, W. Sandhas, J. Haidenbauer, and A. Nogga, {\it Few-Body Systems} {\bf 28}, 241 (2000).

\bibitem{Haidenbauer-PRC53} J. Haidenbauer and K.
Holinde, {\it Phys. Rev.} {\bf C53}, R26 (1995).

\bibitem{Schadow-NPA631} W. Schadow and W. Sandhas, {\it Nucl. Phys. } {\bf A631}, 588c (1998).

\bibitem{Schadow-PRC63} W. Schadow, O. Nohadani, and W. Sandhas, {\it Phys. Rev.} {\bf C63}, 044006 (2001).

\bibitem{Adam-PRC69} J. Adam, Jr., M. T. Pe\~{n}a, and A. Stadler, {\it Phys. Rev.} {\bf C69}, 034008 (2004).

\bibitem{Glockle-PRL71} W. Gl\"{o}ckle and H. Kamada, {\it Phys. Rev. Lett.} {\bf 71}, 971 (1993).

\bibitem{Li-PRC45} G. Q. Li, R. Machleidt, and R. Brockmann, {\it Phys. Rev.} {\bf C45}, 2782 (1992).

\bibitem{Witala-PRC43} H. Witala, W. Gl\"{o}ckle and H. Kamada, {\it Phys. Rev.} {\bf C43}, 1619 (1991).

\bibitem{Glockle-Springer83} W. Gl\"{o}ckle, {\it The Quantum Mechanical Few-Body Problem.}
(Springer-Verlag, Berlin, 1983).

\bibitem{Wolfenstein-PR96} L. Wolfenstein, {\it Phys. Rev.} {\bf 96}, 1654 (1954).

\bibitem{Hadizadeh-preparation} M. R. Hadizadeh and S. Bayegan, {\it in preparation}.



\end{thebibliography}
\end{document}